\documentclass[british]{aa}  
\usepackage{natbib}
\usepackage[british]{babel}           
\usepackage{graphicx}                 
\usepackage{longtable}                
\usepackage{subfig}
\usepackage{caption}
\usepackage{everysel}
\usepackage{keyval}
\usepackage{ragged2e}
\bibpunct{(}{)}{;}{a}{}{,}

\newcommand{\pct}{phase coherent timing}
\newcommand{\pcts}{phase coherent timing solution}
\newcommand{\pctss}{phase coherent timing solutions}
\newcommand{\xmm}{XMM-Newton}

\newcommand{\ltsima}{$\buildrel < \over \sim$}
\newcommand{\lsim}{\lower.5ex\hbox{\ltsima}}
\newcommand{\gtsima}{$\buildrel > \over \sim$}
\newcommand{\gsim}{\lower.5ex\hbox{\gtsima}}

\newcommand{\rxj}{\hbox{\object{RX\,J0720.4$-$3125}}}

\newcommand{\rxf}{\hbox{\object{RX\,J1856.4$-$3754}}}

\begin{document}
\bibliographystyle{aa}
   \title{Updated phase coherent timing solution of the isolated neutron star \rxj\ using 
          recent XMM-Newton and Chandra observations 
	  \thanks{Based on observations with XMM-Newton, an ESA Science Mission with 
	          instruments and contributions directly funded by ESA Member states and the USA (NASA)}}
   
\titlerunning{Timing solution of \rxj}
   \subtitle{}

   \author{M.M. Hohle\inst{1,2} \and F. Haberl\inst{1} 
   															\and J. Vink\inst{3}
                                \and R. Turolla\inst{4,5}                                       
                                \and S. Zane\inst{5}
                                \and C.P. de Vries\inst{6}
                                \and M. M\'endez\inst{7}      
                                }
\authorrunning{Hohle et al.}
   \offprints{Markus M. Hohle, mhohle@astro.uni-jena.de}

   \institute{Max-Planck-Institut f\"ur extraterrestrische Physik, Giessenbachstra{\ss}e, 85741 Garching, Germany
\and Astrophysikalisches Institut und Universit\"ats-Sternwarte Jena, Schillerg\"asschen 2-3, 07745 Jena, Germany 
\and University Utrecht, PO Box 80000, 3508 TA Utrecht, The Netherlands
\and Department of Physics, University of Padua, via Marzolo 8, 35131 Padova, Italy 
\and Mullard Space Science Laboratory, University College London, Holmbury St. Mary, Dorking, Surrey, RH5, 6NT, UK
\and SRON, Netherlands Institute of Space Research, Sorbonnelaan 2, 3584 CA, Utrecht, The Netherlands
\and Kapteyn Astronomical Institute, University of Groningen, P.O. Box 800, 9700 AV Groningen, The Netherlands
}             


 
  \abstract
{}
{Since the last \pcts\ of the nearby radio-quiet isolated neutron star \rxj\ six new \xmm\ and three Chandra observations were carried out. The \pctss\ from previous authors were performed without restricting to a fixed energy band. However, we recently showed that the phase residuals are energy dependent, and thus phase coherent solutions must be computed referring always to the same energy band.}
{We updated the \pcts\ for \rxj\ by including the recent \xmm\ EPIC-pn, MOS1, MOS2 and Chandra ACIS data in the energy range 400-1000~eV. Altogether these observations cover a time span of almost 10~yrs. A further timing solution was obtained including the ROSAT pointed data. In this case, observations cover a time span of $\approx$16~yrs. To illustrate the timing differences between the soft band (120-400~eV) and the hard band (400-1000~eV) a timing solution for the soft band is also presented and the results are verified using a $\mathrm{Z_{n}^{2}}$ test.}
{In contrast to previous work, we obtain almost identical solutions whether or not we include the ROSAT or Chandra data. Thanks to the restriction to the hard band, the data points from EPIC-pn are in better agreement with those from MOS1, MOS2 and Chandra than in previous works. In general the phase residuals are still large and vary with time. In particular, the latest \xmm\ and Chandra data show that the phase residuals have attained relatively large and negative values. Using this and previous timing solutions, the residuals indicate a cyclic behaviour with a period $\mathrm{\approx7-9~yrs}$ if the variations follow a sinusoid, or twice this value in case the residuals are modulated by an abs(sine) probably approaching a new minimum around MJD=55000~days (September 2009). As an alternative interpretation, the phase residuals can be fitted with a glitch that occured around MJD=53000~days.
}
{}
 

\maketitle
%

\section{Introduction}

Since the discovery of the first isolated radio-quiet neutron star \rxf\ \citep{1996Natur.379..233W} in the ROSAT All-Sky Survey only six more soft X-ray sources with similar properties were identified. They are often referred to as the ``Magnificent Seven" (hereafter M7; two more objects, which may be M7-like were found by \citealt{2008ApJ...672.1137R} and \citealt{2009A&A...498..233P}).
Unlike other isolated neutron stars the M7 exhibit pure thermal (black body) emission (with absorption features in some cases) with temperatures $\leq$100~eV (see \citealt{2007Ap&SS.308..181H}). They are connected to faint blue optical counterparts (see \citealt{2008AIPC..968..129K} for a review) with $\mathrm{m_{B}\approx25-28}$. Although these objects are dim in the optical bands, the optical fluxes are $\approx5-10$ times larger than expected by extrapolating the black body X-ray spectrum at low energies (for RBS~1774 the optical excess is a factor of $\approx30$, see \citealt{2008ApJ...682..487Z}). 

A definite explanation of the emission properties of the M7 has not been put forward as yet. Their nearly Planckian spectrum at X-ray energies, in fact, appears difficult to reconcile with the predictions of standard atmospheric models. It has been suggested that their surface layers are in a condensed rather than gaseous state. The phase transition is expected to occur for low temperatures (T$\lsim10^{6}$~K) and high magnetic field strengths (B$\gsim10^{13}$~G; see \citealt{2004ApJ...603..265T} and references therein and \citealt{2006PhRvA..74f2507M,2006PhRvA..74f2508M,2007MNRAS.382.1833M}). If a thin hydrogen atmosphere covers the star, condensed surface models may also explain the optical excess \citep{2007inss.conf.....Z, 2004AdSpR..33..531Z, 2007MNRAS.375..821H}. 

Among the M7, \rxj\ (discovered by \citealt{1997A&A...326..662H} and identified with a faint blue optical star in
\citealt{1998A&A...333L..59M,1998ApJ...507L..49K,2003ApJ...590.1008K,2010AN....331..243E}) has a unique place inasmuch it shows significant variations in its surface temperature, in the equivalent width of an absorption feature seen in the X-ray spectrum and in the size of the emitting area (see \citealt{2004A&A...415L..31D, 2007Ap&SS.308..181H} and \citealt{2009A&A...498..811H}; hereafter H09). 
  
The constant spin down $\dot{P}$ of \rxj\ was first estimated to be $\approx10^{-14}$~s/s in \citet{2002MNRAS.334..345Z} and an updated timing solution using further observations was presented in \citet{2004MNRAS.351.1099C}. The \pctss\ in \citet{2005ApJ...628L..45K} and \citet[][hereafter KvK05 and vK07, respectively]{2007ApJ...659L.149V}, have been computed by including either data from ROSAT, Chandra and \xmm\ (``all data" solution), or Chandra data only and all data except ROSAT (see vK07). Applying a constant spin-down model leads to large phase residuals, therefore, vK07 included a glitch event in one of their timing solutions which significantly reduced the phase residuals.\\
These data sets are from various instruments with different energy responses and in different data acquisition modes. However, \citet{2001A&A...365L.302C} found a hardness ratio variation and a phase shift between the flux and the hardness ratio (both folded into the 8.391~s pulse period) using \xmm{} data. Later, \citet{2004A&A...415L..31D} showed that the energy-dependent change in the pulse profile is accompanied by a long-term change of the X-ray spectrum and proposed precession as a possible explanation. \citet{2004A&A...419.1077H} confirmed the dependence of the pulse profiles on the energy and reported a phase lag between soft (120-400~eV) and hard (400-1000~eV) photons. A long term period of $\mathrm{\approx7}$~yrs was found in the spectral variations and the phase residuals, supporting the precession model \citep{2006A&A...451L..17H}. The phase lag exhibits a long term evolution, its presence appears not to depend on the template used to fit the X-ray pulse profile and it changed sign around MJD=52800 days (H09). This behaviour has to be taken into account and therefore requires timing solutions for both bands separately. Since the \xmm\ MOS and the Chandra HRC and ACIS detectors are less sensitive in the soft band, we focus on a \pcts\ for the hard band only. Since the last \pcts\ in vK07, eight new \xmm{} and twelve new Chandra observations were performed, thus an updated timing solution including the new data is required to distinguish between a periodic trend or a single glitch event.

\section{Observations and Data reduction}

\addtocounter{table}{1} 

We use EPIC (pn, \citealt{2001A&A...365L..18S}, MOS1 and MOS2, \citealt{2001A&A...365L..27T}) data from all 14 \xmm\ observations performed between May 2000 and November 2007 (for a summary of the instrumental setups see H09) and include the most recent \xmm\ observations of March 2009 (rev. 1700, performed in full frame mode with the thin filter for pn and both MOS) and September 2009 (rev. 1792, performed in full frame mode with the thin filter for pn and in small window mode with thin filter for both MOS). We analysed the \xmm\ data with the standard \xmm\ Science Analysis System (SAS) version 7.1.0. applying barycentric correction. We use single and double events for both MOS and the pn detector, that are collected within a circular region of 30" (pn) and 8" (MOS) radius. The observations of March and September 2009 were analysed with SAS version 9.0.0, but following the same procedure as for the other observations. All observations are filtered using the GTI files that are generated during the standard data reduction process. The \xmm\ observation of March 2009 was strongly affected by flares and the GTI files are not sufficient enough to filter out all time intervals with contaminated counts, i.e. for pn only the second half of the exposure time could be used (10.8~ks with $\approx30\times10^{3}$ photons in the hard band), while the MOS data are much less affected. This cut of the pn data is justified posteriori, since the results (periods, phase residuals etc.) for pn and MOS1 \& MOS2 are in good agreement, as it will be seen later.

The Chandra data were analysed with CIAO 4.1. For best corrections of the read-out times and dither in the Advanced CCD Imaging Spectrometer (ACIS, \citealt{2003SPIE.4851...28G}) Continuous Clocking (CC) data, we first checked that the observations were processed with the Standard Data Processing (SDP) version DS 7.6.3. or later. The coordinate accuracy for all ACIS-CC observations is better than 0.5", i.e. one pixel. We applied the Charge Transfer Inefficiency (CTI) correction and use the source photons from a rectangular region of 1"$\times$1" size covering the brightest pixels (located on chip seven for all ACIS-CC observations). For High-Resolution Camera (HRC, \citealt{1997SPIE.3114...53K}) data we used photons from zeroth order (circular region of 2" radius) and both first orders within the standard LETG spectral extraction windows (but limited to $\mathrm{10-60~\AA}$ and the brightest 1" pixel stripes only). Finally we applied the barycentric correction using \texttt{axbary} for the HRC and ACIS-CC data.    

The ROSAT data were taken from \citet{2004MNRAS.351.1099C} whom we refer to for details of the observations and the data reduction.

Assuming that no significant variations in the period of \rxj\ occur within a few days, we merged several close Chandra observations with the same instrumental setup to reduce the scatter in the phase folded light curves. After this merging we obtain 70 data sets: 16 from \xmm\ EPIC-pn and 36 from EPIC-MOS1/MOS2 (the \xmm\ observations rev. 0622 and 0711 from May and October 2003, respectively, have four MOS data sets each), 6 ROSAT data sets (pointings only) and 12 (out of 29) data sets from Chandra observations. All observations are listed in Tab.~\ref{chandraobs}, first column.

\section{The selection of the hard band (400-1000eV)}

\begin{figure}
\centering
{
\includegraphics*[viewport=95 0 440 720, width=0.48\textwidth]{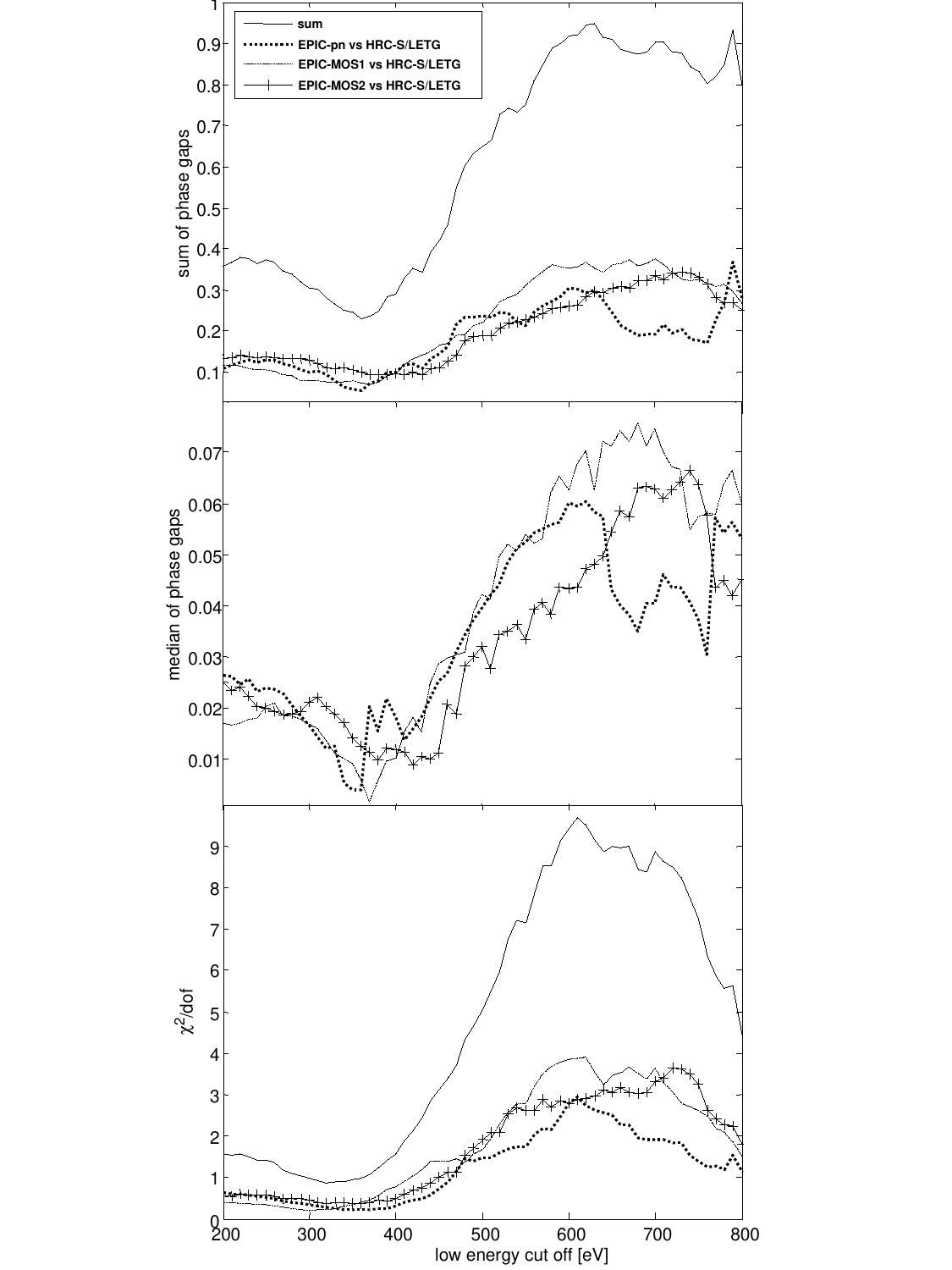}
}
\caption{The accordance of the phase residuals of the hard band (cut off energy$-$1000~eV) for \xmm\ EPIC-pn, EPIC-MOS1 \& MOS2 and Chandra HRC of \rxj\ for different values of the energy band cut. Data sets taken with different observatories were selected in such a way that they are as close in time as possible. The upper panel shows the sum of the absolute values of the phase differences (phase gaps) derived from the different instruments with respect to each other, while the middle panel shows the median of these phase gaps and the corresponding $\mathrm{\chi^{2}/d.o.f.}$ is presented in the lower panel.}
\label{bands}%
\end{figure}

\begin{figure}
\centering
{
\includegraphics*[viewport=115 0 450 720, width=0.48\textwidth]{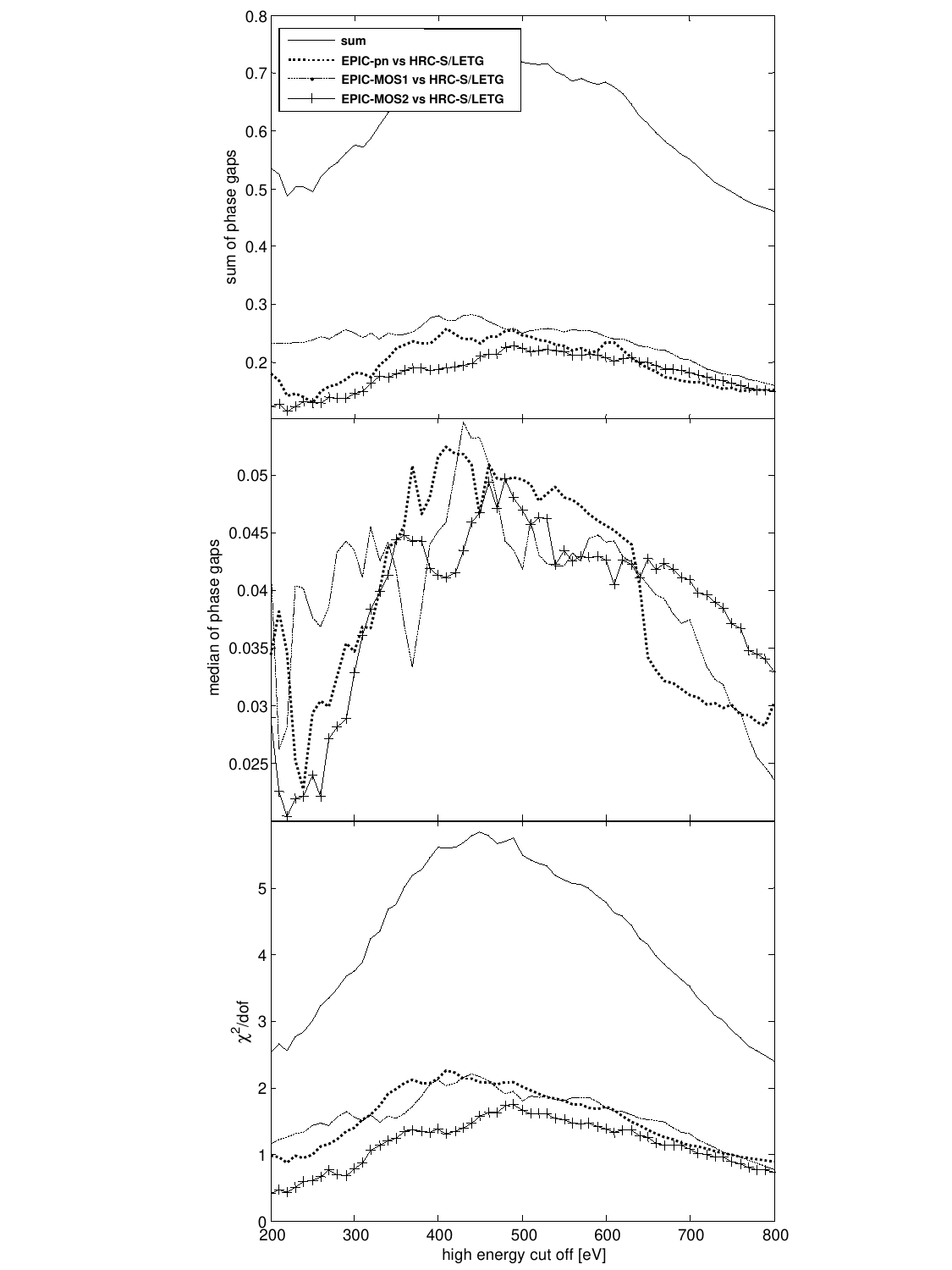}
}
\caption{The same as in Fig~\ref{bands}, but using the soft band (120~eV$-$cut off energy) for the analysis.}
\label{bands_soft}%
\end{figure} 

In previous work \citep[][and H09]{2004A&A...419.1077H,2006A&A...451L..17H} a variable phase lag between soft (120-400~eV) and hard (400-1000~eV) photons of \rxj{} was reported and discussed. Depending on the different detectors and acquisition modes, the data contains different fractions of photons from the two bands that lead to a systematic scatter in the phase residuals. This phenomenon has to be taken into account and requires a different treatment as for the other M7 if an adequate timing solution has to be found. 

In this section we first discuss the cross calibration issues for the different instruments used for the observations of \rxj{}, then derive an empirical energy band that minimises the scatter in the residuals and discuss the results.

In \citet{2004A&A...419.1077H,2006A&A...451L..17H} the photons were divided into a soft (120-400~eV) and a hard band (400-1000~eV) for the EPIC-pn detector to illustrate the phase shift between these two bands. The effective area of the MOS detectors deviates much from that of the pn detector (particularly in the soft band) and the spectral resolution of the MOS detectors is $\sim$~100~eV in the soft band \citep{2001A&A...365L..27T}. This resolution is much worse than for gratings of \xmm{} or Chandra. In addition, the CCD detectors suffer under the contamination of hard photons in the soft band due to redistribution. The MOS and pn observations were executed in different filters: thin, medium and thick. The thick filter significantly suppresses the penetration of soft photons compared to e.g. the thin filter, that also influences the timing properties.

The phase residuals from the \xmm\ observations should be comparable to those of the Chandra observations, taking different energy response, spectral resolution and the phase shift between hard and soft photons into account (softer photons are less prominent in the MOS, HRC and ACIS data than in the pn, influencing the residuals as shown in H09, Fig. 1 and \citealt{2007Ap&SS.308..181H} Fig. 7). The HRC observations have a low number of photons and the energy information of the photons in the zeroth order is lost. Therefore we included all photons in the HRC data. If we would exclude the zeroth order photons, $\mathrm{\sim40\%}$ of the counts were lost and if only the hard band (first orders) would be used, $\mathrm{\sim80\%}$ of the counts were lost, resulting too low statistics. For similar reasons (low statistics, low energy resolution) we also used ROSAT data from a wider energy band: channels 10 to 90 (100-900~eV) for the Position Sensitive Proportional Counter (PSPC) and channels one to eight for the High Resolution Imager (HRI, \citealt{dhk96}).

The energy band minimising the residuals can be found using those Chandra observations performed close to \xmm\ observations: The HRC observation 5582 (June 1, 2005) is close to the \xmm\ observation rev. 0986 (April 28, 2005 ), the HRC observations 6369 \& 7177 (Oct 8/9, 2005) are close to the \xmm\ observation rev. 1060 (Sep 23, 2005), the HRC observations 7243$-$7245 (Dec 14-17, 2005) are close to the \xmm\ observation rev. 1086 (Nov 11/12, 2005), the HRC observations 10861 \& 10700 (Jan 20 \& Feb 14, 2009) are close to the \xmm\ observation rev. 1700 (Mar 21, 2009) and the HRC observation 10701 (Sep 11, 2009) is close to the \xmm\ observation rev. 1792 (Sep 22, 2009) see also Tab.~\ref{chandraobs}, column 6 and 7. Unfortunately no ACIS observation is close in time to these observations.

Assuming that the timing properties of \rxj\ do not change significantly within a time span of a month, we compare the timing residuals (using the ``all data" timing solution in vK07) of these observations directly changing the band cut from 200~eV to 800~eV in steps of 10~eV. For example, if the band cut is at 600~eV, we have two bands of 120-600~eV and 600-1000~eV. Due to the limited energy resolution, these band cuts could not be applied for the HRC observations.

We then calculate the phase residuals by fitting a sine for the phase folded light curves (see vK07) in 12 phase bins for pn, MOS1 and MOS2 and HRC for each band cut and calculate the sum of the absolute values and the median of the phase gaps of the phase residuals of the harder band from the different detectors. Having the phase residuals and their errors, we also calculate $\chi^2$/d.o.f showing the degree of accordance of the results from the different detectors. In all three cases, we derive a minimum (i.e. best agreement) if we use the band cut between 300~eV and 400~eV (for consistency with previous work, \citealt[][and H09]{2004A&A...419.1077H,2006A&A...451L..17H}, we set the cut between the two bands at 400~eV), i.e. having two bands of 120-400~eV and 400-1000~eV and using the harder band (see Figure~\ref{bands}). For band cuts of lower energies the different detector responses for the soft photons cause larger phase gaps, because of the phase shift between hard and soft photons and the rough energy resolution of the EPIC detectors, while for band cuts at higher energies the number of photons decreases and the statistics worsens.

If the same is done, but keeping the soft band photons (see Fig.~\ref{bands_soft}), the best agreement of the different instruments is achieved if no band cut is applied. But even the lowest $\chi^2$/d.o.f value is more than two times larger as in the case when photons between 400-1000~eV are used.
 
Apart from the intrinsic properties of \rxj{} (the spectrum did not change much since rev. 0986, see H09), the choice of this band cut for the hard photons at least reduces instrumental discrepancies of the different detectors.

\section{Timing procedure}

For comparison with the results of KvK05 (obtained using the $\mathrm{Z_{n}^{2}}$ method, \citealt{1983A&A...128..245B}), we start taking as reference period that from the ACIS-CC observations 4666-4669 combined with the HRC observation 5305, covering a total time span of 52 days. We determine the period from the peak in the $\mathrm{Z_{1}^{2}}$ periodogram (from now on always $\mathrm{n=1}$) and estimated the 1$\sigma$ errors following \citet{2002ApJ...570L..79K} using the equation derived in \citet{2002AJ....124.1788R}. The period we have taken corresponds to the center of the 1$\sigma$ region in the periodogram (since the peak is symmetric close to its maximum). 

Using this dataset produces aliases in the periodogram caused by the time gaps between the observations. We obtain three peaks with $\mathrm{Z_{1}^{2}\approx140}$: $P_{1}=8.39036664(53)$~s, $P_{2}=8.39111600(50)$~s and $P_{3}=8.39188114(53)$~s. $P_{2}$ yields formally the largest $\mathrm{Z_{1}^{2}}$ value and is consistent with the reference period $P=8.39111590(50)$~s from KvK05 (numbers in parentheses indicate 1$\sigma$ errors using the error estimation in \citealt{2002AJ....124.1788R}). During the timing procedure we found $P_{2}$ being the correct period since starting with $P_{1}$ or $P_{3}$ as initial period, the procedure does not converge.

As a first step we determined the period of each observation separately using the $\mathrm{Z_{1}^{2}}$ method. The period of the Chandra~HRC observation 7251 shows a large discrepancy with respect to the others, probably caused by the small number of photons giving not enough statistics for a reliable period determination. Therefore we excluded this observation from our analysis. The periods obtained from the ROSAT data with a low count number differ sensibly from the reference period and deviate significantly from those periods listed in \citet{2004MNRAS.351.1099C} and are highly uncertain, thus are also excluded for the first \pct\ solution. No periodic signal was found in the ROSAT all sky survey, i.e. this observation is not used at all and not listed in Tab.~\ref{chandraobs}. 

Most of the individual periods with more than $\mathrm{10^{4}}$ photons are consistent with the single spin-down model within 1$\sigma$ errors \citep{2002AJ....124.1788R} and all of them are consistent within 2$\sigma$ errors, both for the hard and the soft band. Note that the periods for Chandra HRC and ROSAT data are the same for both bands since we applied no energy selection for these observations, as mentioned before. All individual periods are listed in Tab.~\ref{chandraobs2}, column 4 and 5. 

\begin{figure}
\centering
\resizebox{\hsize}{!}
{
\includegraphics[viewport=85 255 490 590, width=0.48\textwidth]{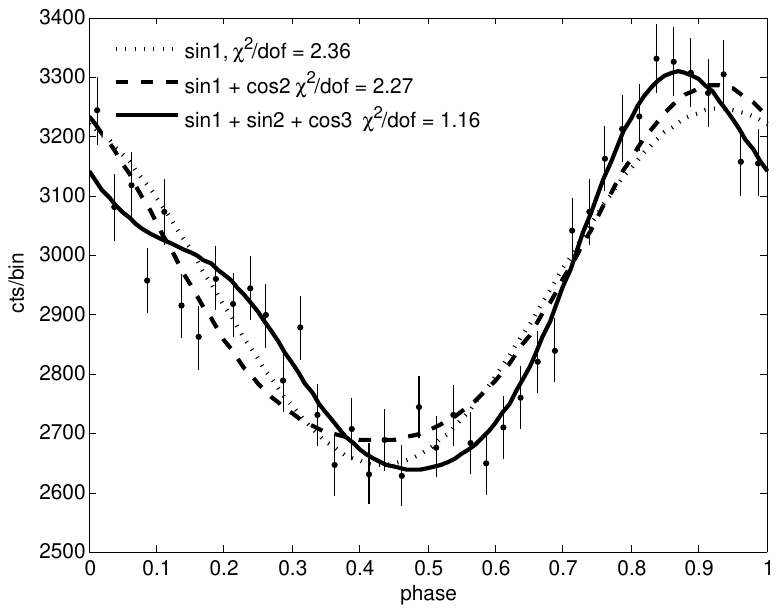}
}
\caption{Phase binned (40 bins) light curve (400-1000~eV) of the \xmm\ EPIC-pn observation revolution 1086 (November 2005) folded during the timing procedure (described in the text). The solid line represents the combination of three Fourier harmonics chosen for the determination of the phase. The phase shift differs slightly from that from the sine fit (dotted line). The dashed line shows, that a fit with the second harmonic (cosine) is not always sufficient. The error bars denote Poissonian errors.}
\label{var}%
\end{figure}

\begin{figure}
\centering
\resizebox{\hsize}{!}
{
\includegraphics*[viewport=40 65 470 680, width=0.48\textwidth]{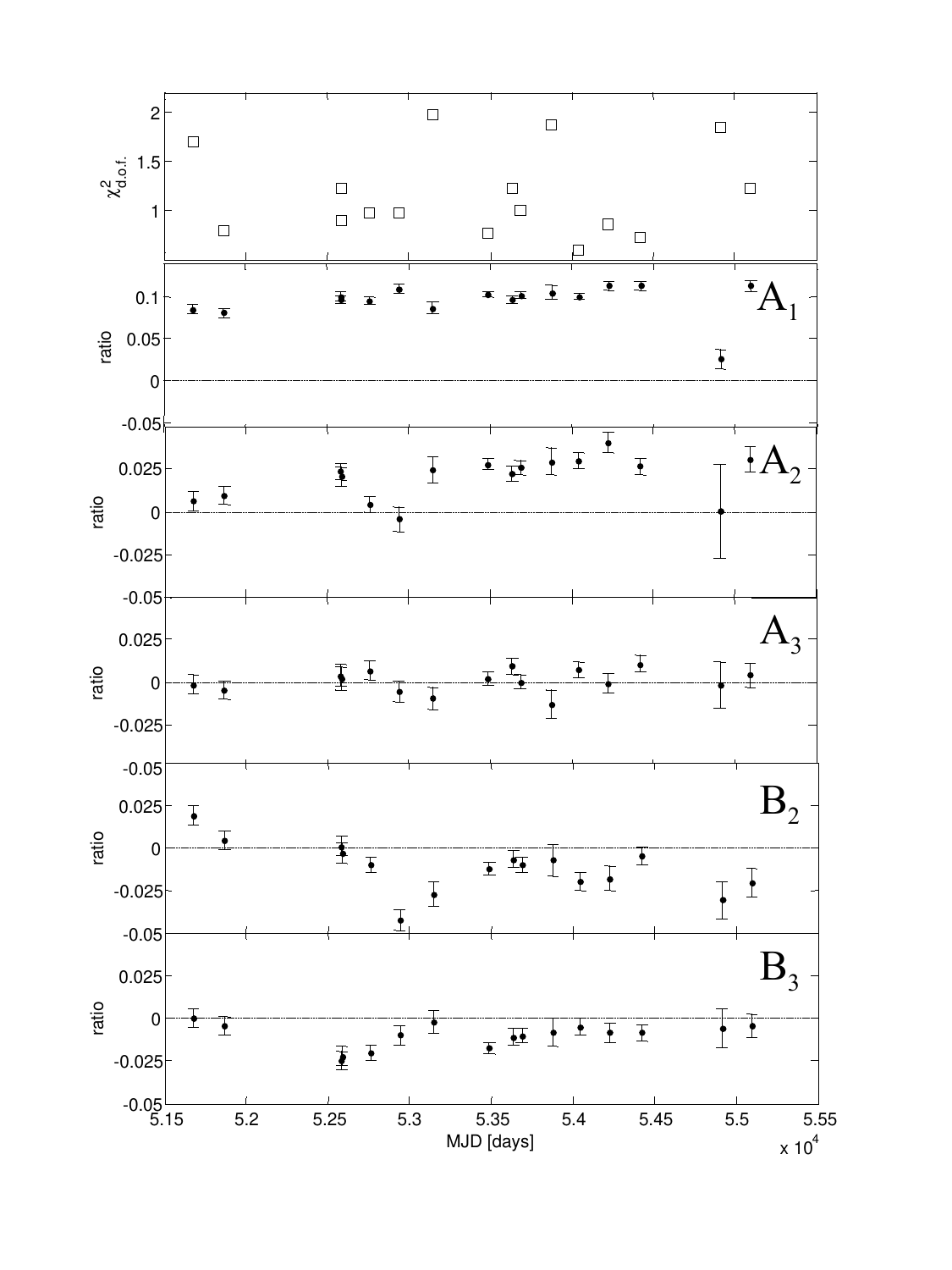}
}
\caption{The Fourier coefficients (cos1=0 by definition) of the fit to the phase folded light curves (40 bins) of the EPIC-pn observations. The values of the coefficients are divided by the constant offset $\mathrm{A_{0}}$, since the observations have different numbers of photons. The first panel shows the corresponding values of $\chi^2/d.o.f.$. All observations have less than 1\% background counts and are filtered using standard GTI files. In case of rev. 1700 ($\mathrm{\sim43\%}$ high background) an additional filtering is applied (see Sec. 2) leading to the deviating point near MJD=54900~days. All error bars denote 1$\sigma$ confidence level.}
\label{fourier_panel}%
\end{figure}

\begin{figure}
\centering
\resizebox{\hsize}{!}
{
\includegraphics*[viewport=40 50 490 715, width=0.48\textwidth]{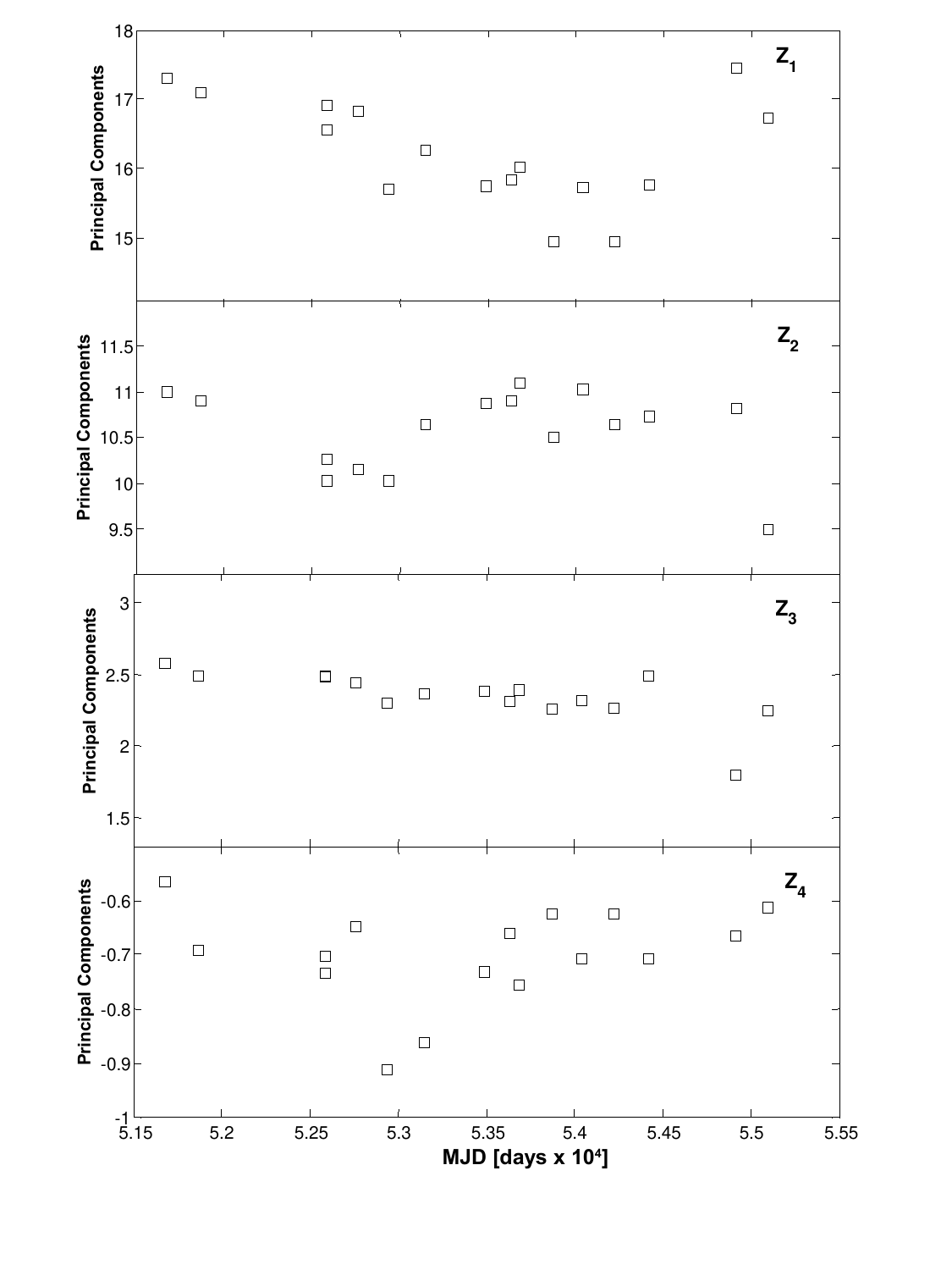}
}
\caption{The principal components (see \citealt{2006MNRAS.366..727Z} for details) of the 16 EPIC-pn observations of \rxj{}.}
\label{PC_panel}%
\end{figure}

Our timing procedure follows that in KvK05 with the difference that we calculate the times of arrival (TOA, the time of maximum light closest to the middle of the observation, see KvK05) from the included observations during the iterations of the timing procedure. The errors of $\dot{f}$ and $P$ are evaluated after each step of the iteration: in the case of $\dot{f}$ the error is calculated from the different $\dot{f}$ values satisfying the condition to minimise the phase residuals within their $\mathrm{1\sigma}$ errors. This restricts the possible values for $\dot{f}$ if more observations are added, i.e. the error on $\dot{f}$ decreases with growing time $T$ from the reference point. The error of $P$, $eP$, automatically decreases with growing time span $T$ following $eP\sim1/T$ (e.g. see the general approach in \citealt{1981Ap&SS..78..175K}).

In KvK05, vK07 and H09 the phase binned light curves were fitted with a sinusoid to derive the phase of the maximum, although the pulse profiles are not always best represented by a sine function \citep{2004A&A...419.1077H}. We fitted the light curves using a combination of three harmonics of the Fourier series\footnote{The light curves are fitted using the $\mathrm{MATLAB^{\textcircled{R}}}$ (version R2008a) internal function {\it fit} with default adjustments. Errors derived from this function and the errors derived from the results of {\it fit} are scaled to $\sqrt{\chi^{2}/dof}$.} as follows: The Fourier series for the light curves is given by
 
\begin{equation}
      \mathrm{F(\phi)=\sum\limits_{k=0}^{\infty} A_{k}sin(k\phi)  + B_{k}cos(k\phi)   }  \,
\label{fourier}     
\end{equation}

where $\mathrm{k=0}$ corresponds to the constant offset. We choose to fit the light curves with the sum of three terms, not to introduce too many degrees of freedom. To simplify the notation, in the following we denote each term just as sink or cosk, e.g. sin1 stands for $\mathrm{A_{1}sin(1\phi)}$, cos1 for $\mathrm{B_{1}cos(1\phi)}$. Since the contribution of cos1 turned out to be negligible (since we forced the phase to be zero at the reference time, and a cosine term would imply a phase shift), our sum starts always with sin1, while the remaining two terms can be sin2+sin3, cos2+sin3 or cos2+sin2, and so on. To keep the number of free parameters small, we do not use terms with $\mathrm{k>3}$ and use only three terms in total. The combination that fits the individual light curve with the lowest value of $\chi^2$/d.o.f. was used for the phase determination. This leads to an improvement of the $\chi^2$/d.o.f. for the light curves (see example in Fig. \ref{var}) and to a better determination of the phase shift. Our approach is justified since in most of the cases some Fourier coefficients are negligible (see the coefficients from the phase folded light curves of the EPIC-pn observations in Fig.~\ref{fourier_panel} as an example).  

The analysis of the Fourier coefficients shows that the pulse profile changes
in time, as already reported in \citet{2006A&A...451L..17H} and \citet[][their Fig. 2]{2004A&A...415L..31D}, and the evolution
continued also after the last observation considered by \citet[][MJD=53700~days]{2006A&A...451L..17H}. In
order to check that the light curve evolution is independent of the adopted
pulse profile template (either a sine as in \citealt{2006A&A...451L..17H}, or a truncated Fourier
expansion, as in the present work), we performed a principal component
analysis (PCA, see e.g. \citealt{2006MNRAS.366..727Z} for more details) of the same 16 EPIC-pn
light curves we used before, again binned in 40 phase intervals. Starting from
the original variables (the 40 values of the counts at the different
phases), the PCA computes a new set of variables (the principal components,
pcs) which are a linear combination of the old ones and are ordered in such a
way that the first pc accounts for the largest variance of the data, the
second the second largest, and so on. We find that the first four pcs are
actually responsible for $\mathrm{\sim 92\%}$ of the variance, pulse profiles can be effectively classified in terms of only a limited number of pcs which embody their main characteristics. The time evolution of the first four pcs
is shown in Fig.~\ref{PC_panel}. The significant changes in both sets of coefficients (pc and Fourier) is a strong evidence of a
genuine variation of the pulse profile in the hard band in time. The second last \xmm{} observation (rev. 1700) that was contaminated by strong flares deviates from the trends in Fig.~\ref{fourier_panel} and Fig.~\ref{PC_panel}.  
As discussed in \citet{2006MNRAS.366..727Z}, the first pc ($\mathrm{Z_{1}}$) is related to the pulse amplitude, $\mathrm{Z_{2}}$ provide a measure of the phase interval of the light maximum and $\mathrm{Z_{3}}$ reflects the lightcurve parity with respect to the half period (see Fig.~4 of
\citealt{2006MNRAS.366..727Z}, for the dependence on the phase of the first four coefficients $v_{ik}$, with $i=1,\ldots, 4$, used to calculate the PCs from the original variables). Although a detailed analysis will not be attempted here, Fig.~\ref{PC_panel} shows that the largest changes occur in $\mathrm{Z_{1}}$ and $\mathrm{Z_{2}}$ and are then associated to the amplitude and the position of the maximum of the pulses.

\begin{figure*}
\centering
\resizebox{\hsize}{!}
{
\includegraphics*[viewport=40 0 680 540, width=0.48\textwidth]{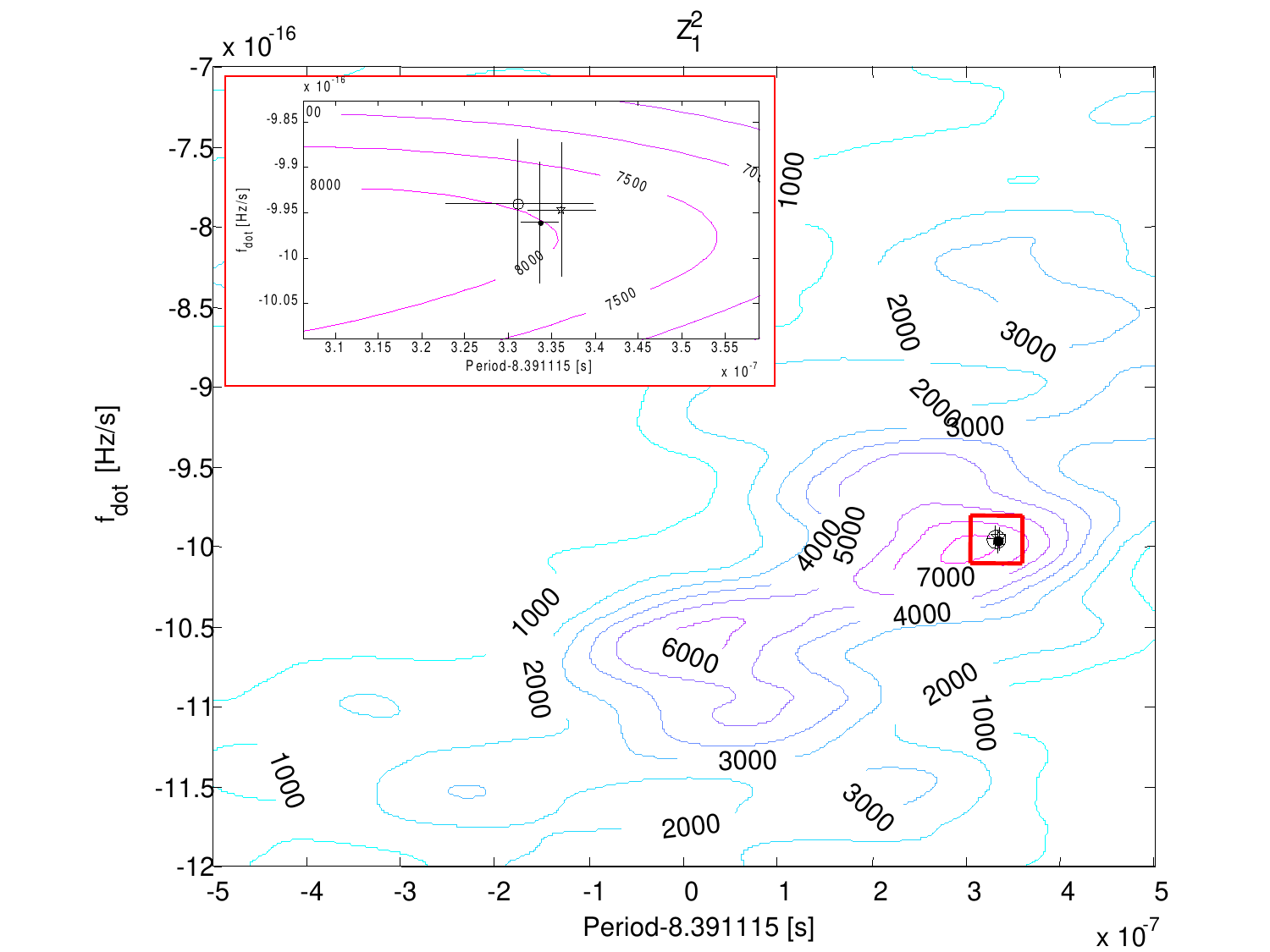}
}
\caption{The $\mathrm{Z_{1}^{2}}$ values in the $P-\dot{f}$ plane derived from the \xmm{}, ROSAT and Chandra observations (except the HRC observation 7251) of \rxj{} compared to the results of the \pctss\ in this work. The peak is located at $P=8.391115309(14)$~s and $\dot{f}=-9.992(61)\times10^{-16}$~Hz/s. The three different timing solutions derived without the Chandra data (open circle), without ROSAT data (star) and using \xmm{}, ROSAT and Chandra observations (dot) are not resolved on this scale (red box), but become visible enlarging the red box (shown in the upper left panel); the values are listed in Tab.~\ref{zusammenfassung}. The errors of the $\mathrm{Z_{1}^{2}}$ solution are scaled to $\sqrt{\chi^{2}/dof}$. All timing solutions are derived from the hard band (400-1000~eV).}
\label{zsquare}%
\end{figure*}

\section{Results}

\addtocounter{table}{1}

We performed the \pct\ by binning the light curves into 10, 12, 16, 18 and 20 phase bins. A large number of bins leads to large scatter in the Chandra light curves due to the small count rate compared to \xmm\, while too few bins result in an insufficient time resolution\footnote{Note that the time resolution for the MOS detectors is 0.3~s (small window), 0.9~s (large window) and 2.6~s (full frame), i.e. less than one phase bin. This is compensated by the large number of photons. KvK05 used 16 phase bins (including the MOS detectors).}. We derived the best timing solution for all observations (excluding ROSAT data and the HRC observation 7251) for 12 phase bins in order to minimise the phase residuals. 

If we fit the phase folded light curves with a pure sinusoid, we obtain $P=8.3911153307(22)$~s, $\dot{f}=-9.933(52)\times10^{-16}$~Hz/s for the \pct\ solution. This corresponds to $\chi^2/d.o.f.=12.7$ for the timing solution and for the light curve fits we obtain $\chi^2/d.o.f.=2.29$ on average. Fitting the light curves with a variable combination of the three Fourier harmonics, the result does not change significantly. 
We obtain $P=8.3911153362(39)$~s and $\dot{f}=-9.946(74)\times10^{-16}$~Hz/s, but the phase residuals have smaller errors and therefore the timing solution corresponds to $\chi^2/d.o.f.=47$. For the light curve fits we obtain $\chi^2/d.o.f.=1.23$ on average.

Although the periods determined from the ROSAT data (see \citealt{2002MNRAS.334..345Z} and \citealt{2004MNRAS.351.1099C} for details) differ significantly to the reference period $P_{2}$, we include these data sets, except the All-Sky Survey (where we find no period), for a further timing solution. The inclusion of the ROSAT data extends the time span of observations from $\approx$9.6~yrs to $\approx$16.5~yrs and enlarges the data set from 64 to 70 observations. Again, we fitted the light curves with a variable combination of the three Fourier harmonics and obtain $P=8.3911153336(22)$~s and $\dot{f}=-9.961(67)\times10^{-16}$~Hz/s for 12 phase bins. This corresponds to $\chi^2/d.o.f.=45$ for the timing solution and for the light curve fits we obtain $\chi^2/d.o.f.=1.17$ on average. 

For a third timing solution we excluded the Chandra data, but using the ROSAT data (58 data sets out of 70) and the same reference time and period as for the other two timing solutions. Fitting the light curves with a variable combination of the three Fourier harmonics, we obtain $P=8.3911153310(22)$~s and $\dot{f}=-9.940(71)\times10^{-16}$~Hz/s for 12 phase bins that corresponds to $\chi^2/d.o.f.=47$ for the timing solution and for the light curve fits we obtain $\chi^2/d.o.f.=1.21$ on average.

To verify the correctness of our timing solutions we apply the $\mathrm{Z_{1}^{2}}$ method to the same combination of observations as for the \pctss\ (see Fig. \ref{zsquare}). Using all 70 data sets, we found a maximum at $P=8.391115309(14)$~s and $\dot{f}=-9.992(61)\times10^{-16}$~Hz/s with $\mathrm{Z_{1}^{2}=8263}$. The errors for $P$ and $\dot{f}$ are obtained from the equation in \citet{2002AJ....124.1788R}, whereas the values correspond to the maximum peak in the center of the 1$\sigma$ confidence region. In this region the peak is symmetric. Finally, the errors are scaled to $\sqrt{\chi^{2}/dof}$.

The results of the three \pctss\ and from the $\mathrm{Z_{1}^{2}}$ method are summarised in Tab.~\ref{zusammenfassung}. Note that our results are only slightly different from previous solutions (but scatter less), i.e. the restriction of our investigations to the hard band reduces the phase residuals and influences the shape of their evolution in time, but does not affect basic parameters like the spin-down age or magnetic field strength of \rxj\ significantly. For completeness the results from the coherent ''all data" solution and the $\mathrm{Z_{1}^{2}}$ method for the soft photons are listed in Tab.~\ref{zusammenfassung} too, illustrating the difference of the timing solutions in the different bands. The corresponding TOAs from the final solutions are listed in Tab.~\ref{chandraobs2}, last two columns. The errors of the TOAs are derived from the formal fit errors of the phase binned light curves by fitting them to the combination of Fourier harmonics as explained in Section 4.

\begin{figure}
\centering
\resizebox{\hsize}{!}
{
\includegraphics[viewport=85 255 490 590, width=0.48\textwidth]{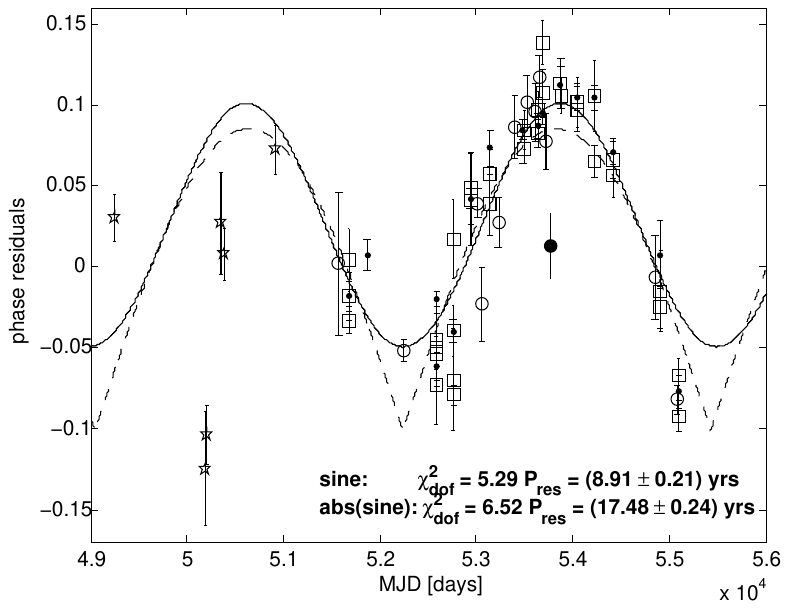}
}
\caption{Phase residuals after applying the timing solutions derived without the ROSAT observations and without the Chandra HRC observation 7251 (marked as a filled circle) from \rxj{} applied to all observations listed in Tab.~\ref{chandraobs}. The best fits with a sine and an abs(sine) model (only the observations used in the current timing solution are used for the fits) are also shown. Squares mark residuals derived from \xmm\ EPIC-MOS1 and MOS2 observations, dots EPIC-pn observations, circles Chandra HRC and ACIS-CC data and stars ROSAT data. All error bars denote 1$\sigma$.}
\label{respanel}%
\end{figure}

\begin{table*}
   \begin{center}
     \caption{$P$ and $\dot{f}$ derived from different methods in this paper, compared to previous results.}
        \label{zusammenfassung}

\begin{tabular}{lllccc}

\hline
\hline

solution&period-8.391115&$\dot{f}$ & rms & d.o.f &  $\chi^2/d.o.f$\\
	&$\times10^{-7}$~[s]&$\times10^{-16}$~[Hz/s] & [s] & & \\
\hline
\multicolumn{6}{c}{this work, hard band} \\
\multicolumn{6}{c}{}\\
``all data" & 3.336(22)& -9.961(67) & 0.62 & 70-3 & 45 \\
\multicolumn{3}{l}{after fitting a sine} & 0.29 & 70-5 & 7.8\\
\multicolumn{3}{l}{after fitting an abs(sine)} &  0.31 & 70-5 & 8.6\\
without ROSAT & 3.362(39)& -9.946(74) & 0.60 & 64-3 & 47 \\
\multicolumn{3}{l}{after fitting a sine} & 0.19 & 64-5 & 5.3\\
\multicolumn{3}{l}{after fitting an abs(sine)} & 0.21 & 64-5 & 6.5\\
without Chandra & 3.310(22)& -9.940(71) & 0.61 & 58-3 & 47 \\
\multicolumn{3}{l}{after fitting a sine} & 0.31 & 58-5 & 9.7\\
\multicolumn{3}{l}{after fitting an abs(sine)} & 0.33 & 58-5 & 8.7\\
$Z_{1}^{2}$ ``all data" &   3.09(14)& -9.992(61) & 0.73 & 70-3 & 64 \\
$Z_{1}^{2}$ without ROSAT & 2.96(13)& -10.047(34) & 0.58 & 64-3 & 46 \\
$Z_{1}^{2}$ without Chandra & 3.09(13)& -9.980(36) & 0.72 & 58-3 & 51 \\
\hline
\multicolumn{6}{c}{this work, soft band} \\
\multicolumn{6}{c}{}\\
``all data" &  3.429(22)  &  -9.956(72) & 0.50 & 70-3 & 40 \\
 $Z_{1}^{2}$ ``all data"  & 3.31(62) & -9.959(17) & 0.50 & 70-3 & 39 \\
\hline
\multicolumn{6}{c}{previous work (applied to the hard band)} \\
\multicolumn{6}{c}{}\\
vK07 (``all data")& 2.670(84) & -9.88(13) & 0.97 & 70-3 & 81 \\
vK07 (without ROSAT)& 2.846(77) & -9.74(04) & 1.30 & 64-3 & 207 \\
KvK05 (``all data")& 3.20(13) & -9.918(15) & 0.64 & 70-3 & 56 \\
KvK05 (Chandra)& 3.05(16) & -9.97(06) & 0.64 & 12-3 & 57 \\
\hline
\end{tabular}
\end{center}
Note: For the timing solution in this work we always excluded the Chandra HRC observation 7251. Since the phase residuals seem to follow a periodic pattern we fitted a sine and an abs(sine), see Sec. 6. All errors correspond to 1$\sigma$ confidence (for the $Z_{1}^{2}$ solution see \citealt{2002AJ....124.1788R} with errors scaled to $\sqrt{\chi^{2}/dof}$, for the phase coherent solutions, see Sec. 5).
\end{table*} 

\section{Discussion}

We present new updated \pctss\ for the isolated radio-quiet X-ray pulsar \rxj\ including the most recent \xmm\ and Chandra observations, restricting our analysis to the hard energy band (400-1000~eV), except for Chandra HRC and ROSAT data. The new solutions were obtained with and without the inclusion of the ROSAT and Chandra data and differ slightly from previous ones (see KvK05, and vK07 and Tab.~\ref{zusammenfassung}). We improved the phase determination and the modeling of the phase folded light curves by fitting a variable combination of the three Fourier harmonics with the sine as the leading term, instead of a pure sinusoid. We checked the \pctss\ applying the $\mathrm{Z_{1}^{2}}$ method to the observations in the $P$/$\dot{f}$ plane. All timing solutions correspond to a $\chi^2/d.o.f.\approx50$, i.e. the phase residuals are still large and a timing solution based only on $\dot{f}$ and $P$ provides an inadequate model for the spin behaviour of \rxj{}. 

Due to the restriction to the hard band for most of the data, the phase residuals from EPIC-pn do not deviate systematically to those from EPIC-MOS1 and MOS2 and Chandra unlike those derived without energy restriction (see Fig. 1 in H09 for comparison and Fig.~\ref{bands} in this work) and most observations are in good agreement to each other. However, the residual of the Chandra observation 7251 is still not consistent (lowest number of counts and bad statistics) to the other phase residuals and the phase residuals from the \xmm\ EPIC-pn and EPIC-MOS1 \& MOS2  observations 0622 scatter much. The observations of rev 0622 were all performed in different filters: thick, medium and thin, having different transparency for the softer photons (causing deviations in the phase as discussed before) in the hard band. 

Including the most recent Chandra and \xmm\ observations, the ROSAT data become less important for the timing solution. This explains why our three \pctss\ are almost identical and scatter less than the \pctss\ in KvK05 and vK07 (without a glitch).

Applying the timing solutions to the observations of \rxj{} we still obtain large residuals. Due to the small number of counts, the ROSAT data are ambiguous in determining the correct phase, the periods differ significantly from the periods of the other observations and the ROSAT periodograms are noisy. However, the inclusion of the ROSAT data extends the observed time span from 9.6 to 16 yrs. Therefore, we present the phase residuals from all observations, but derived from the timing solution without ROSAT data (hard band) in Fig.~\ref{respanel} shown with an error weighted sine and an abs(sine) fit (see also H09). The phase residuals reached a minimum around MJD$=$52800~days and continue towards negative values, maybe a new minimum around MJD$=$55000~days; but this has to be confirmed by new observations finding a possible turning point. There is no clear evidence for a minimum in the past (probably MJD$=$50000~days including the ROSAT data), but the ROSAT data are less reliable.

Fitting a sine or an abs(sine) to the phase residuals reduces the $\chi^2/d.o.f.$ from $\approx50$ to $\approx6-7$ with a period $\mathrm{P_{res}\approx8.9}$~yrs\footnote{In H09 long term periods of $\mathrm{P_{res}\approx5.5-7.5}$~yrs were found for the sine fit. This is significantly less than found in this work. However, the timing solutions used in H09 were not derived from the inclusion of the new data but were adopted from KvK05 and vK07, also required the inclusion of an additional slope (not necessary in this work).} (i.e. the observations cover two cycles) or twice this period fitting an abs(sine), respectively. However, the $\chi^2/d.o.f.$ is formally still unacceptable. A sinusoidal behaviour of the phase residuals could be explained by precession, as suggested in \citet{2004A&A...415L..31D} and \citet{2006A&A...451L..17H}. Precession would cause an advanced (the NS precesses towards the observer) and a retarded (the NS precesses backwards with respect to the observer) signal, whereas the residuals would follow a sine \citep{1990ApJ...348..226N}. During precession the observer would see different parts of the surface that would lead to changing spectral properties, such as changes in temperature and size of the emitting surface. If two, roughly antipodal, hot spots would both contribute to the X-ray emission, the phase residuals caused from both hot spots would be shifted by $\mathrm{P_{res}/2}$: if one hot spot precesses towards the observer and appears, the second hot spot moves backwards with respect to the observer and disappears. Since the timing is not sensitive to which of the hot spots is causing the residuals, the phase residuals qualitatively may have the shape of an abs(sine), having twice the period of the corresponding sine (two peaks from two different hot spots). The presence of the larger period is supported by the variable phase lags between hard and soft photons, see \citet[][and H09]{2004A&A...419.1077H, 2006A&A...451L..17H}. However, it has to be shown whether this scenario can explain the spectral behaviour in H09 too. Formally, then also the spectral changes undergo a periodic behaviour having $\mathrm{P_{res}\approx2\times8.9}$~yrs. A long term period of $\mathrm{P_{res}\approx17-18}$~yrs could explain why the spectral changes do not show a periodic trend (H09) yet, since the time span of the \xmm{} observations would cover less than one cycle. The new turning point of the phase residuals would be at $\mathrm{MJD=55387^{+107}_{-100}~days}$ (first half of July 2010) or at $\mathrm{MJD=55240^{+8}_{-8}~days}$ (first half of February 2010) for the sine or the abs(sine) fit, respectively (see Fig.~\ref{respanel}).   

Alternatively, vK07 proposed a glitch around MJD=53000~days explaining the spectral and temporal changes. As shown in H09, the temperature increased at this time, followed by a slow decrease, as expected during and after a glitch event. Therefore vK07 presented a ``glitch solution" to explain also the timing behaviour of \rxj{}. This glitch solution was not applied correctly in H09. Indeed, the glitch solution presented in vK07 minimises the phase residuals for the observations available at that time (MJD=53500~days), i.e. the statement (in H09) that it produces large residuals is not correct. However, adding new observations after MJD=53500~days (not available in vK07) and applying the glitch solution correctly, the phase residuals grow to larger values. This can be shown even with the observations available in H09 (MJD=54421~days) and is even more pronounced using the recent data (until MJD=55100~days), i.e. the conclusions in H09 remains the same.

To reduce the residuals, we introduce a new additional parameter with the physical meaning of a post-glitch correction for $\dot{f}$. This can be done by fitting a parabola with respect to the time distance to the proposed glitch time, $\mathrm{t_{g}=52866~days}$, in vK07 (``all data" solution). The error weighted fit for the hard band including all data gives $\dot{f}_{c}=-1.11(10)\times10^{-17}$~Hz/s and significantly reduces the phase residuals ($\chi^2/d.o.f.=2.8$, rms=0.31~s). vK07 obtained $\dot{f}_{vK07}=-1.04(3)\times10^{-15}$~Hz/s after $\mathrm{t_{g}}$, the new value would be $\dot{f}_{new}=\dot{f}_{vK07}+\dot{f}_{c}$, i.e. a modification of the glitch solution in vK07. Due to this correction the glitch hypothesis is still a competing model. 

\begin{figure}
\centering
\resizebox{\hsize}{!}
{
\includegraphics*[viewport=78 130 460 665, width=0.48\textwidth]{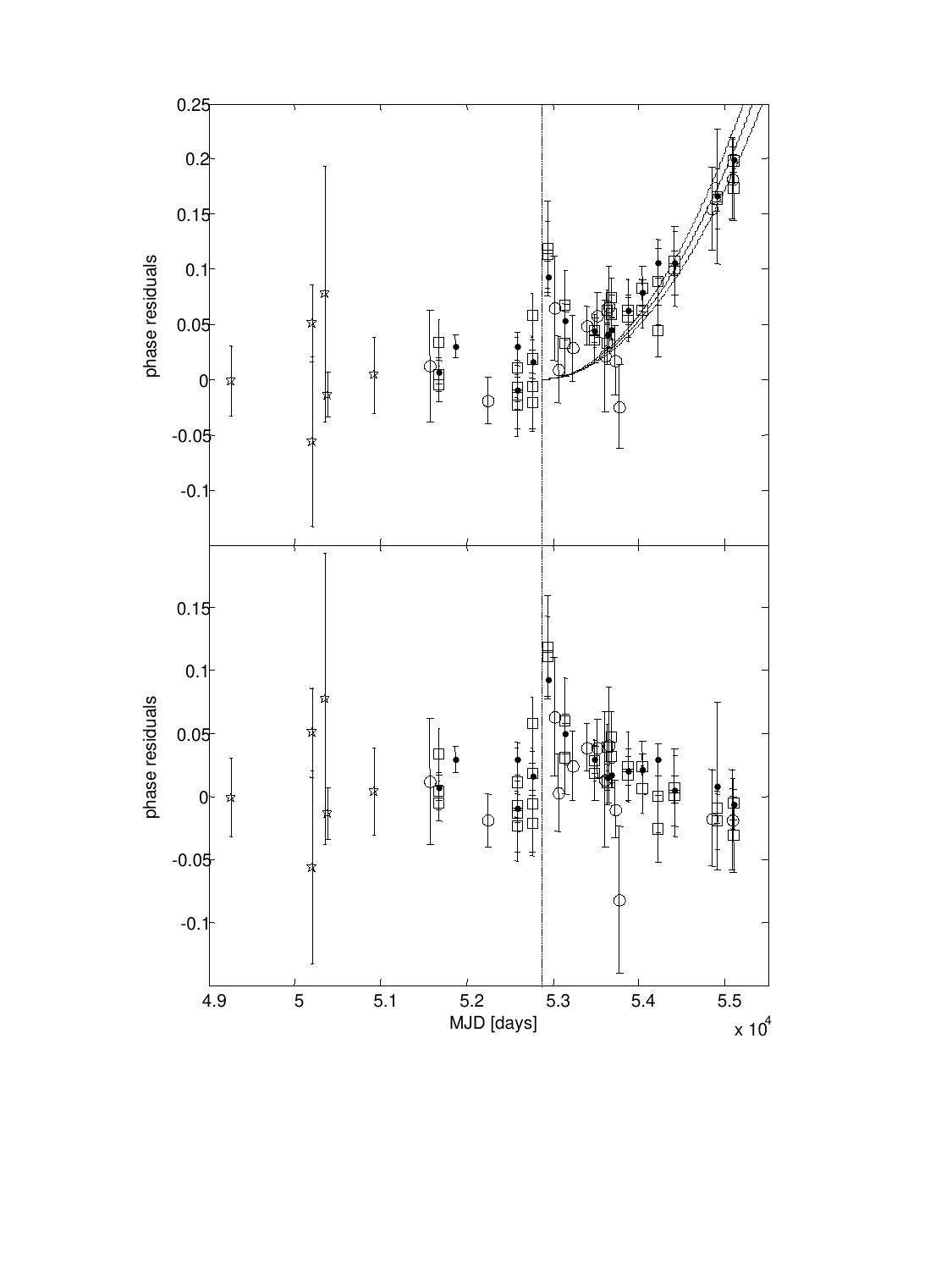}
}
\caption{Upper panel: The phase residuals of \rxj\ after applying the glitch solution (``all data") in vK07. The glitch time $\mathrm{t_{g}=52866~days}$ is marked as dotted vertical line. We fitted an error wighted parabolic slope (solid line, dashed lines indicating the 1$\sigma$ confidence range.) corresponding to a post-glitch correction of $\dot{f}_{c}=-1.11(10)\times10^{-17}$~Hz/s. Lower panel: The phase residuals after applying the post-glitch $\dot{f}$ correction. All symbols like in Fig.~\ref{respanel}}
\label{glitch}%
\end{figure}

The phase residuals of \rxj\ after applying the glitch solution in vK07 and the post-glitch correction (both for the hard band) are shown in Fig.~\ref{glitch}. 

Recently, an analysis of timing irregularities of 366 radio pulsars \citep{2010MNRAS.402.1027H} showed that quasi$-$periodic structures in the timing residuals are quiet usual - at least for radio pulsars. These quasi$-$periodic structures are dominant for young pulsars (age$\mathrm{\leq10^{5}~yrs}$, note that \rxj\ is probably much younger than implied by its characteristic age, see \citealt{2010MNRAS.402.2369T,2007ApJ...660.1428K}). In some cases the amplitude of the timing residuals are comparable to those of \rxj\ and the timing residuals show ''periods" of a few years. These radio pulsars were studied over decades and the data points are much denser in time than in the case of X-ray pulsars, thus the timing residuals of \rxj\ may also follow such trends. In such cases (as the authors argue in \citealt{2010MNRAS.402.1027H}) the timing residuals caused due to ''slow glitches" may not be a different phenomenon to that causing the timing irregularities.

The behaviour of \rxj\ is still not understood. The monitoring of \rxj\ using \xmm\ and Chandra is still ongoing and will help to bring us closer to understand the reason for its behaviour.

\begin{acknowledgements}
The \xmm\ project is supported by the Bundesministerium f\"ur Wirtschaft und
Technologie/Deutsches Zentrum f\"ur Luft- und Raumfahrt (BMWI/DLR, FKZ 50 OX 0001)
and the Max-Planck Society.
MMH acknowledges support by the Deutsche Forschungsgemeinschaft (DFG) through
SFB/TR 7 ``Gravitationswellenastronomie'' and CompStar, a research networking programme of the European Science Foundation (ESF).
The work of RT is partially funded by INAF-ASI through grant AAE TH-058.
We like to thank the referee Marten van Kerkwijk for the fast and detailed review of the manuscript and for useful comments which helped in improving the paper.
\end{acknowledgements}

\bibliography{ins,general,myrefereed,myunrefereed}

\begin{thebibliography}{41}
\expandafter\ifx\csname natexlab\endcsname\relax\def\natexlab#1{#1}\fi

\bibitem[{{Buccheri} {et~al.}(1983){Buccheri}, {Bennett}, {Bignami}, {Bloemen},
  {Boriakoff}, {Caraveo}, {Hermsen}, {Kanbach}, {Manchester}, {Masnou},
  {Mayer-Hasselwander}, {Ozel}, {Paul}, {Sacco}, {Scarsi}, \&
  {Strong}}]{1983A&A...128..245B}
{Buccheri}, R., {Bennett}, K., {Bignami}, G.~F., {et~al.} 1983, \aap, 128, 245

\bibitem[{{Cropper} {et~al.}(2004){Cropper}, {Haberl}, {Zane}, \&
  {Zavlin}}]{2004MNRAS.351.1099C}
{Cropper}, M., {Haberl}, F., {Zane}, S., \& {Zavlin}, V.~E. 2004, \mnras, 351,
  1099

\bibitem[{{Cropper} {et~al.}(2001){Cropper}, {Zane}, {Ramsay}, {Haberl}, \&
  {Motch}}]{2001A&A...365L.302C}
{Cropper}, M., {Zane}, S., {Ramsay}, G., {Haberl}, F., \& {Motch}, C. 2001,
  \aap, 365, L302

\bibitem[{{David} {et~al.}(1996){David}, {Harnden}, {Kearns}, \&
  {Zombeck}}]{dhk96}
{David}, L., {Harnden}, J.~F., {Kearns}, K., \& {Zombeck}, M. 1996, The ROSAT
  High Resolution Imager (HRI), USRSDC/SAO Calibration Report, revised

\bibitem[{{de Vries} {et~al.}(2004){de Vries}, {Vink}, {M{\'e}ndez}, \&
  {Verbunt}}]{2004A&A...415L..31D}
{de Vries}, C.~P., {Vink}, J., {M{\'e}ndez}, M., \& {Verbunt}, F. 2004, \aap,
  415, L31

\bibitem[{{Eisenbeiss} {et~al.}(2010){Eisenbeiss}, {Ginski}, {Hohle},
  {Hambaryan}, {Neuh{\"a}user}, \& {Schmidt}}]{2010AN....331..243E}
{Eisenbeiss}, T., {Ginski}, C., {Hohle}, M.~M., {et~al.} 2010, Astronomische
  Nachrichten, 331, 243

\bibitem[{{Garmire} {et~al.}(2003){Garmire}, {Bautz}, {Ford}, {Nousek}, \&
  {Ricker}}]{2003SPIE.4851...28G}
{Garmire}, G.~P., {Bautz}, M.~W., {Ford}, P.~G., {Nousek}, J.~A., \& {Ricker},
  Jr., G.~R. 2003, in Society of Photo-Optical Instrumentation Engineers (SPIE)
  Conference Series, Vol. 4851, Society of Photo-Optical Instrumentation
  Engineers (SPIE) Conference Series, ed. J.~E. {Truemper} \& H.~D.
  {Tananbaum}, 28--44

\bibitem[{{Haberl}(2007)}]{2007Ap&SS.308..181H}
{Haberl}, F. 2007, \apss, 308, 181

\bibitem[{{Haberl} {et~al.}(1997){Haberl}, {Motch}, {Buckley}, {Zickgraf}, \&
  {Pietsch}}]{1997A&A...326..662H}
{Haberl}, F., {Motch}, C., {Buckley}, D. A.~H., {Zickgraf}, F.~J., \&
  {Pietsch}, W. 1997, \aap, 326, 662

\bibitem[{{Haberl} {et~al.}(2006){Haberl}, {Turolla}, {de Vries}, {Zane},
  {Vink}, {M{\'e}ndez}, \& {Verbunt}}]{2006A&A...451L..17H}
{Haberl}, F., {Turolla}, R., {de Vries}, C.~P., {et~al.} 2006, \aap, 451, L17

\bibitem[{{Haberl} {et~al.}(2004){Haberl}, {Zavlin}, {Tr{\"u}mper}, \&
  {Burwitz}}]{2004A&A...419.1077H}
{Haberl}, F., {Zavlin}, V.~E., {Tr{\"u}mper}, J., \& {Burwitz}, V. 2004, \aap,
  419, 1077

\bibitem[{{Ho} {et~al.}(2007){Ho}, {Kaplan}, {Chang}, {van Adelsberg}, \&
  {Potekhin}}]{2007MNRAS.375..821H}
{Ho}, W.~C.~G., {Kaplan}, D.~L., {Chang}, P., {van Adelsberg}, M., \&
  {Potekhin}, A.~Y. 2007, \mnras, 375, 821

\bibitem[{{Hobbs} {et~al.}(2010){Hobbs}, {Lyne}, \&
  {Kramer}}]{2010MNRAS.402.1027H}
{Hobbs}, G., {Lyne}, A.~G., \& {Kramer}, M. 2010, \mnras, 402, 1027

\bibitem[{{Hohle} {et~al.}(2009){Hohle}, {Haberl}, {Vink}, {Turolla},
  {Hambaryan}, {Zane}, {de Vries}, \& {M{\'e}ndez}}]{2009A&A...498..811H}
{Hohle}, M.~M., {Haberl}, F., {Vink}, J., {et~al.} 2009, \aap, 498, 811(H09)

\bibitem[{{Kaplan}(2008)}]{2008AIPC..968..129K}
{Kaplan}, D.~L. 2008, in American Institute of Physics Conference Series, Vol.
  968, Astrophysics of Compact Objects, ed. {Y.-F.~Yuan, X.-D.~Li, \& D.~Lai},
  129--136

\bibitem[{{Kaplan} {et~al.}(2002){Kaplan}, {Kulkarni}, {van Kerkwijk}, \&
  {Marshall}}]{2002ApJ...570L..79K}
{Kaplan}, D.~L., {Kulkarni}, S.~R., {van Kerkwijk}, M.~H., \& {Marshall}, H.~L.
  2002, \apjl, 570, L79

\bibitem[{{Kaplan} \& {van Kerkwijk}(2005)}]{2005ApJ...628L..45K}
{Kaplan}, D.~L. \& {van Kerkwijk}, M.~H. 2005, \apjl, 628, L45(KvK05)

\bibitem[{{Kaplan} {et~al.}(2007){Kaplan}, {van Kerkwijk}, \&
  {Anderson}}]{2007ApJ...660.1428K}
{Kaplan}, D.~L., {van Kerkwijk}, M.~H., \& {Anderson}, J. 2007, \apj, 660, 1428

\bibitem[{{Kaplan} {et~al.}(2003){Kaplan}, {van Kerkwijk}, {Marshall},
  {Jacoby}, {Kulkarni}, \& {Frail}}]{2003ApJ...590.1008K}
{Kaplan}, D.~L., {van Kerkwijk}, M.~H., {Marshall}, H.~L., {et~al.} 2003, \apj,
  590, 1008

\bibitem[{{Kovacs}(1981)}]{1981Ap&SS..78..175K}
{Kovacs}, G. 1981, \apss, 78, 175

\bibitem[{{Kraft} {et~al.}(1997){Kraft}, {Chappell}, {Kenter}, {Kobayashi},
  {Meehan}, {Murray}, {Zombeck}, {Fraser}, {Pearson}, {Lees}, {Brunton},
  {Barbera}, {Collura}, \& {Serio}}]{1997SPIE.3114...53K}
{Kraft}, R.~P., {Chappell}, J.~H., {Kenter}, A.~T., {et~al.} 1997, in Society
  of Photo-Optical Instrumentation Engineers (SPIE) Conference Series, Vol.
  3114, Society of Photo-Optical Instrumentation Engineers (SPIE) Conference
  Series, ed. O.~H. {Siegmund} \& M.~A. {Gummin}, 53--73

\bibitem[{{Kulkarni} \& {van Kerkwijk}(1998)}]{1998ApJ...507L..49K}
{Kulkarni}, S.~R. \& {van Kerkwijk}, M.~H. 1998, \apjl, 507, L49

\bibitem[{{Medin} \& {Lai}(2006{\natexlab{a}})}]{2006PhRvA..74f2507M}
{Medin}, Z. \& {Lai}, D. 2006{\natexlab{a}}, \pra, 74, 062507

\bibitem[{{Medin} \& {Lai}(2006{\natexlab{b}})}]{2006PhRvA..74f2508M}
{Medin}, Z. \& {Lai}, D. 2006{\natexlab{b}}, \pra, 74, 062508

\bibitem[{{Medin} \& {Lai}(2007)}]{2007MNRAS.382.1833M}
{Medin}, Z. \& {Lai}, D. 2007, \mnras, 382, 1833

\bibitem[{{Motch} \& {Haberl}(1998)}]{1998A&A...333L..59M}
{Motch}, C. \& {Haberl}, F. 1998, \aap, 333, L59

\bibitem[{{Nelson} {et~al.}(1990){Nelson}, {Finn}, \&
  {Wasserman}}]{1990ApJ...348..226N}
{Nelson}, R.~W., {Finn}, L.~S., \& {Wasserman}, I. 1990, \apj, 348, 226

\bibitem[{{Pires} {et~al.}(2009){Pires}, {Motch}, {Turolla}, {Treves}, \&
  {Popov}}]{2009A&A...498..233P}
{Pires}, A.~M., {Motch}, C., {Turolla}, R., {Treves}, A., \& {Popov}, S.~B.
  2009, \aap, 498, 233

\bibitem[{{Ransom} {et~al.}(2002){Ransom}, {Eikenberry}, \&
  {Middleditch}}]{2002AJ....124.1788R}
{Ransom}, S.~M., {Eikenberry}, S.~S., \& {Middleditch}, J. 2002, \aj, 124, 1788

\bibitem[{{Rutledge} {et~al.}(2008){Rutledge}, {Fox}, \&
  {Shevchuk}}]{2008ApJ...672.1137R}
{Rutledge}, R.~E., {Fox}, D.~B., \& {Shevchuk}, A.~H. 2008, \apj, 672, 1137

\bibitem[{{Str{\"u}der} {et~al.}(2001){Str{\"u}der}, {Briel}, {Dennerl},
  {Hartmann}, {Kendziorra}, {Meidinger}, {Pfeffermann}, {Reppin}, {Aschenbach},
  {Bornemann}, {Br{\"a}uninger}, {Burkert}, {Elender}, {Freyberg}, {Haberl},
  {Hartner}, {Heuschmann}, {Hippmann}, {Kastelic}, {Kemmer}, {Kettenring},
  {Kink}, {Krause}, {M{\"u}ller}, {Oppitz}, {Pietsch}, {Popp}, {Predehl},
  {Read}, {Stephan}, {St{\"o}tter}, {Tr{\"u}mper}, {Holl}, {Kemmer}, {Soltau},
  {St{\"o}tter}, {Weber}, {Weichert}, {von Zanthier}, {Carathanassis}, {Lutz},
  {Richter}, {Solc}, {B{\"o}ttcher}, {Kuster}, {Staubert}, {Abbey}, {Holland},
  {Turner}, {Balasini}, {Bignami}, {La Palombara}, {Villa}, {Buttler},
  {Gianini}, {Lain{\'e}}, {Lumb}, \& {Dhez}}]{2001A&A...365L..18S}
{Str{\"u}der}, L., {Briel}, U., {Dennerl}, K., {et~al.} 2001, \aap, 365, L18

\bibitem[{{Tetzlaff} {et~al.}(2010){Tetzlaff}, {Neuh{\"a}user}, {Hohle}, \&
  {Maciejewski}}]{2010MNRAS.402.2369T}
{Tetzlaff}, N., {Neuh{\"a}user}, R., {Hohle}, M.~M., \& {Maciejewski}, G. 2010,
  \mnras, 402, 2369

\bibitem[{{Turner} {et~al.}(2001){Turner}, {Abbey}, {Arnaud}, {Balasini},
  {Barbera}, {Belsole}, {Bennie}, {Bernard}, {Bignami}, {Boer}, {Briel},
  {Butler}, {Cara}, {Chabaud}, {Cole}, {Collura}, {Conte}, {Cros}, {Denby},
  {Dhez}, {Di Coco}, {Dowson}, {Ferrando}, {Ghizzardi}, {Gianotti}, {Goodall},
  {Gretton}, {Griffiths}, {Hainaut}, {Hochedez}, {Holland}, {Jourdain},
  {Kendziorra}, {Lagostina}, {Laine}, {La Palombara}, {Lortholary}, {Lumb},
  {Marty}, {Molendi}, {Pigot}, {Poindron}, {Pounds}, {Reeves}, {Reppin},
  {Rothenflug}, {Salvetat}, {Sauvageot}, {Schmitt}, {Sembay}, {Short},
  {Spragg}, {Stephen}, {Str{\"u}der}, {Tiengo}, {Trifoglio}, {Tr{\"u}mper},
  {Vercellone}, {Vigroux}, {Villa}, {Ward}, {Whitehead}, \&
  {Zonca}}]{2001A&A...365L..27T}
{Turner}, M. J.~L., {Abbey}, A., {Arnaud}, M., {et~al.} 2001, \aap, 365, L27

\bibitem[{{Turolla} {et~al.}(2004){Turolla}, {Zane}, \&
  {Drake}}]{2004ApJ...603..265T}
{Turolla}, R., {Zane}, S., \& {Drake}, J.~J. 2004, \apj, 603, 265

\bibitem[{{van Kerkwijk} {et~al.}(2007){van Kerkwijk}, {Kaplan}, {Pavlov}, \&
  {Mori}}]{2007ApJ...659L.149V}
{van Kerkwijk}, M.~H., {Kaplan}, D.~L., {Pavlov}, G.~G., \& {Mori}, K. 2007,
  \apjl, 659, L149(vK07)

\bibitem[{{Walter} {et~al.}(1996){Walter}, {Wolk}, \&
  {Neuh{\"a}user}}]{1996Natur.379..233W}
{Walter}, F.~M., {Wolk}, S.~J., \& {Neuh{\"a}user}, R. 1996, \nat, 379, 233

\bibitem[{{Zane} {et~al.}(2002){Zane}, {Haberl}, {Cropper}, {Zavlin}, {Lumb},
  {Sembay}, \& {Motch}}]{2002MNRAS.334..345Z}
{Zane}, S., {Haberl}, F., {Cropper}, M., {et~al.} 2002, \mnras, 334, 345

\bibitem[{{Zane} {et~al.}(2008){Zane}, {Mignani}, {Turolla}, {Treves},
  {Haberl}, {Motch}, {Zampieri}, \& {Cropper}}]{2008ApJ...682..487Z}
{Zane}, S., {Mignani}, R.~P., {Turolla}, R., {et~al.} 2008, \apj, 682, 487

\bibitem[{{Zane} \& {Turolla}(2006)}]{2006MNRAS.366..727Z}
{Zane}, S. \& {Turolla}, R. 2006, \mnras, 366, 727

\bibitem[{{Zane} {et~al.}(2004){Zane}, {Turolla}, \&
  {Drake}}]{2004AdSpR..33..531Z}
{Zane}, S., {Turolla}, R., \& {Drake}, J.~J. 2004, Advances in Space Research,
  33, 531

\bibitem[{{Zane} {et~al.}(2007){Zane}, {Turolla}, \&
  {Page}}]{2007inss.conf.....Z}
{Zane}, S., {Turolla}, R., \& {Page}, D., eds. 2007, {Isolated Neutron Stars:
  from the Surface to the Interior}

\end{thebibliography}

\longtab{1}{
\begin{longtable}{rl|cc|ccr}
\caption{All observations of \rxj\ listed in chronological order.} \label{chandraobs}\\
\hline
Obs. Id.	& Instrument/setup 	 &    Counts              &         Counts                    & MJD   & Start Date  & Effective \\
        	&                  	 &    (120-400~eV)        &         (400-1000~eV)             &       &             & exposure  \\     
		      &                    & 			                  &				                            & [days]&             & [ks]      \\
\hline
\endfirsthead
\caption[]{-- Continued. --}\\\hline
Obs. Id.	& Instrument/setup 	 &    Counts            &         Counts                    & MJD   & Start Date  & Effective \\
        	&                  	 &    (120-400~eV)      &         (400-1000~eV)             &       &             & exposure  \\     
		      &                    & 			                &				                            & [days]&             & [ks]      \\
\hline
\endhead
\hline
\endfoot
\hline
\endlastfoot

rp300338n00 	& PSPC        	 & \multicolumn{2}{c|}{5374}				      & 49258 & 1993 Sep 27 &  3.22 \\
rh300508n00 	& HRI         	 & \multicolumn{2}{c|}{1259}				      & 50199 & 1996 Apr 25 &  3.13 \\  
rh180100n00 	& HRI          	 & \multicolumn{2}{c|}{1197}				      & 50211 & 1996 May  7 &  3.57 \\ 
rh300508n01 	& HRI          	 & \multicolumn{2}{c|}{493}  				      & 50354 & 1996 Sep 27 &  1.41 \\  
rh400884n00 	& HRI          	 & \multicolumn{2}{c|}{13381}				      & 50391 & 1996 Nov  3 & 33.57 \\
h400944n00 	& HRI          	 & \multicolumn{2}{c|}{3054}				      & 50924 & 1998 Apr 20 &  3.57 \\	 
368$^{1}$     	& HRC-S/LETG    	 & \multicolumn{2}{c|}{2722}				      & 51575 & 2000 Feb  1 &  5.40 \\
745$^{1}$     	& HRC-S/LETG    	 & \multicolumn{2}{c|}{9392}				      & 51576 & 2000 Feb  2 & 26.26 \\
369$^{1}$     	& HRC-S/LETG    	 & \multicolumn{2}{c|}{2660}				      & 51579 & 2000 Feb  4 &  6.12 \\
0078 S3	& EPIC-pn/FF thin	 & 241783		  & 115287   			      & 51677 & 2000 May 13 & 58.60 \\  
     S1	& EPIC-MOS1/FF thin	 &  69623		  &  43977			      & 51677 & 2000 May 13 & 61.98 \\  
     S2	& EPIC-MOS2/SW thin 	 &  68010		  &  40818   			      & 51677 & 2000 May 13 & 61.99 \\  
0175 S3	& EPIC-pn/FF med & 103491		  &  55743			      & 51870 & 2000 Nov 21 & 25.65 \\  
     S7	& EPIC-MOS1/LW med &  15765		  &  11662			      & 51870 & 2000 Nov 21 & 18.00 \\  
2774$^{2}$   	& ACIS-CC         	 &  13455		  &  15648			      & 52248 & 2001 Dec  4 & 15.01 \\  
2773$^{2}$   	& ACIS-CC          	 &   9452		  &  11646			      & 52248 & 2001 Dec  5 & 10.61 \\  
2771$^{2}$ 	& ACIS-CC         	 &    876		  &   1553			      & 52250 & 2001 Dec  6 &  1.86 \\
2772$^{2}$ 	& ACIS-CC         	 &   3537		  &   4487			      & 52250 & 2001 Dec  6 &  4.05 \\  
0533 S3	& EPIC-pn/FF thin	 & 137951		  &  72150			      & 52585 & 2002 Nov  6 & 28.38 \\  
     S1	& EPIC-MOS1/FF thin	 &  27987		  &  21892 			      & 52585 & 2002 Nov  6 & 29.99 \\  
     S2	& EPIC-MOS2/FF thin	 &  29169		  &  22192			      & 52585 & 2002 Nov  6 & 29.99 \\  
0534 S3	& EPIC-pn/FF thin	 & 145057		  &  76095			      & 52587 & 2002 Nov  8 & 30.18 \\  
     S1	& EPIC-MOS1/FF thin	 &  28168		  &  22112			      & 52587 & 2002 Nov  8 & 31.80 \\  
     S2	& EPIC-MOS2/FF thin	 &  28954		  &  22768			      & 52587 & 2002 Nov  8 & 31.79 \\  
0622 U2	& EPIC-pn/SW thick	 & 123199		  &  94655			      & 52762 & 2003 May  2 & 72.79 \\  
     S5	& EPIC-MOS1/FF med &  21204		  &  18741			      & 52762 & 2003 May  2 & 29.98 \\  
     S3	& EPIC-MOS1/FF thin	 &  28433		  &  23754			      & 52762 & 2003 May  2 & 33.59 \\  
     S6	& EPIC-MOS2/FF med &  21988		  &  19650			      & 52762 & 2003 May  2 & 29.99 \\  
     S4	& EPIC-MOS2/FF thin	 &  30901		  &  24413			      & 52762 & 2003 May  2 & 33.59 \\  
0711 S8	& EPIC-pn/SW med &  83151		  &  60557			      & 52940 & 2003 Oct 27 & 24.90 \\  
     S3	& EPIC-MOS1/FF thin	 &  10841		  &  11937			      & 52940 & 2003 Oct 27 & 13.90 \\  
     S5	& EPIC-MOS1/LW thin	 &  10547		  &  11769			      & 52940 & 2003 Oct 27 & 15.71 \\  
     S4	& EPIC-MOS2/FF thin	 &  11649		  &  12392			      & 52940 & 2003 Oct 27 & 13.91 \\  
     S6	& EPIC-MOS2/LW thin	 &  11943		  &  12382			      & 52940 & 2003 Oct 27 & 15.71 \\  
4666$^{3}$      & ACIS-CC          	 &   5606		  &  13094			      & 53010 & 2004 Jan  6 & 10.12 \\  
4667$^{3}$      & ACIS-CC         	 &   2678		  &   5835			      & 53011 & 2004 Jan  7 &  4.79 \\  
4668$^{3}$      & ACIS-CC          	 &   2121		  &   4918			      & 53016 & 2004 Jan 11 &  5.16 \\  
4669$^{3}$      & ACIS-CC               &   1890		  &   4277			      & 53023 & 2004 Jan 19 &  5.22 \\  
5305$^{4}$	& HRC-S/LETG            & \multicolumn{2}{c|}{13275}				      & 53062 & 2004 Feb 27 & 35.70 \\  
0815 S1	& EPIC-pn/FF thin   & 130683		  &  93280			      & 53147 & 2004 May 22 & 41.30 \\    
     S2	& EPIC-MOS1/FF thin   &  35853		  &  41178			      & 53147 & 2004 May 22 & 45.21 \\   
     S3	& EPIC-MOS2/FF thin   &  35915		  &  42318			      & 53147 & 2004 May 22 & 45.22 \\   
4670$^{5}$      & ACIS-CC          	 &   5723		  &  12329			      & 53221 & 2004 Aug  3 & 10.13 \\
4671$^{5}$      & ACIS-CC               &   2778		  &   6318			      & 53223 & 2004 Aug  5 &  5.15 \\
4672$^{5}$      & ACIS-CC               &   2483		  &   5789			      & 53227 & 2004 Aug  9 &  5.12 \\  
4673$^{5}$      & ACIS-CC               &   2527		  &   6034			      & 53244 & 2004 Aug 23 &  5.13 \\	
5581$^{6}$    	& HRC-S/LETG            & \multicolumn{2}{c|}{30998}				      & 53393 & 2005 Jan 23 & 68.20 \\    
0986 S3	& EPIC-pn/FF thin	 & 181754		  & 131146			      & 53488 & 2005 Apr 28 & 51.43 \\    
     S1	& EPIC-MOS1/SW thin 	 &  32400		  &  36475			      & 53488 & 2005 Apr 28 & 53.05 \\	      
     S2	& EPIC-MOS2/SW thin 	 &  34456		  &  38266			      & 53488 & 2005 Apr 28 & 53.06 \\	      
5582$^{7}$  	& HRC-S/LETG    	 & \multicolumn{2}{c|}{35777}				      &	53523 & 2005 Jun  1 & 70.17 \\	      
6364$^{8}$	& HRC-S/LETG            & \multicolumn{2}{c|}{22210}				      & 53610 & 2005 Aug 27 & 38.87 \\	      
1060 S3	& EPIC-pn/FF thin	 & 175481		  & 120913			      & 53636 & 2005 Sep 23 & 51.14 \\	      
     S1	& EPIC-MOS1/SW thin 	 &  33710		  &  37070			      & 53636 & 2005 Sep 23 & 52.76 \\	  
     S2	& EPIC-MOS2/SW thin 	 &  33977		  &  37706			      & 53636 & 2005 Sep 23 & 52.77 \\	  
6369$^{9}$    	& HRC-S/LETG            & \multicolumn{2}{c|}{7696}				      &	53652 & 2005 Oct  8 & 26.26 \\
7177$^{9}$    	& HRC-S/LETG            & \multicolumn{2}{c|}{2532}				      & 53653 & 2005 Oct  9 &  8.04 \\	       
1086 S3	& EPIC-pn/FF thin	 & 167725		  & 117828			      & 53687 & 2005 Nov 12 & 37.84 \\	      
     S1	& EPIC-MOS1/SW thin 	 &  25265		  &  27091			      & 53687 & 2005 Nov 12 & 39.46 \\			  
     S2	& EPIC-MOS2/SW thin 	 &  25045		  &  27580			      & 53687 & 2005 Nov 12 & 39.47 \\		  
7243$^{10}$   	& HRC-S/LETG            & \multicolumn{2}{c|}{5931}				      &	53718 & 2005 Dec 14 & 17.18 \\
7244$^{10}$   	& HRC-S/LETG            & \multicolumn{2}{c|}{4927}				      &	53718 & 2005 Dec 15 & 16.29 \\
7245$^{10}$   	& HRC-S/LETG            & \multicolumn{2}{c|}{5249}  				      &	53718 & 2005 Dec 16 & 17.19 \\
5584$^{10}$   	& HRC-S/LETG            & \multicolumn{2}{c|}{5321}				      &	53718 & 2005 Dec 17 & 14.19 \\		  
{\it 7251}	& {\it HRC-S/LETG}            & \multicolumn{2}{c|}{{\it 4787}}				      & {\it 53775} & {\it 2006 Sep 9}  & {\it 10.65} \\    	 	    
1181 S1	& EPIC-pn/FF thin	 &  89015		  &  59524			      &	53877 & 2006 May 22 & 20.04 \\	       
     S2	& EPIC-MOS1/SW thin 	 &  13652		  &  14219			      &	53877 & 2006 May 22 & 21.66 \\		  
     S3	& EPIC-MOS2/SW thin 	 &  15045		  &  15048			      &	53877 & 2006 May 22 & 21.66 \\    
1265 S1	& EPIC-pn/FF thin	 &  89852		  &  60883			      & 54044 & 2006 Nov  5 & 20.04 \\
     S2	& EPIC-MOS1/FF thin	 &  12414		  &  12296			      & 54044 & 2006 Nov  5 & 21.61 \\	 
     S3	& EPIC-MOS2/FF thin	 &  15291		  &  14653   			      & 54044 & 2006 Nov  5 & 21.62 \\	 
1356 S1	& EPIC-pn/FF thin	 &  89411		  &  57483			      & 54226 & 2007 May  5 & 20.04 \\	 
     S2	& EPIC-MOS1/FF thin	 &  16690		  &  16685			      & 54226 & 2007 May  5 & 21.61 \\	 
     S3	& EPIC-MOS2/FF thin	 &  17147		  &  16818			      & 54226 & 2007 May  5 & 21.62 \\   
1454 S1	& EPIC-pn/FF thin	 & 102917		  &  64833			      & 54421 & 2007 Nov 17 & 23.06 \\   
     S2	& EPIC-MOS1/FF thin	 &  18674		  &  18197			      & 54421 & 2007 Nov 17 & 24.62 \\   
     S3	& EPIC-MOS2/FF thin	 &  19029		  &  18416			      & 54421 & 2007 Nov 17 & 24.62 \\   
10861$^{11}$	& HRC-S/LETG            & \multicolumn{2}{c|}{4240}				      &	54851 & 2009 Jan 20 & 11.91 \\
10700$^{11}$   	& HRC-S/LETG            & \multicolumn{2}{c|}{14053}				      &	54876 & 2009 Feb 14 & 21.82 \\		     
1700 S3	& EPIC-pn/FF thin	 &  37085		  &  30806			      & 54913 & 2009 Mar 21 & 10.84 \\		 
     U2	& EPIC-MOS1/FF thin	 &  11481		  &  11876			      & 54912 & 2009 Mar 21 & 17.69 \\		 
     U2	& EPIC-MOS2/FF thin	 &  11936		  &  11556			      & 54912 & 2009 Mar 21 & 17.69 \\
10701$^{12}$   	& HRC-S/LETG            & \multicolumn{2}{c|}{15647}				      & 55086 & 2009 Sep 11 & 33.17 \\		 
1792 S3	& EPIC-pn/FF thin	 &  65407		  &  48105			      & 55096 & 2009 Sep 22 & 17.90 \\		 
     S1	& EPIC-MOS1/SW thin 	 &  12524		  &  11919			      & 55096 & 2009 Sep 22 & 19.22 \\	       
     S2	& EPIC-MOS2/SW thin 	 &  12724		  &  11993			      & 55096 & 2009 Sep 22 & 19.24 \\		
																       
\hline																       
\end{longtable}
\noindent Note: Observation 7251, in {\it italic}, is not used in this work (see Sec. 3). The identifier marks the merged data sets: observations with the same tag are merged into one data set. Counts for ROSAT and Chandra HRC were not divided into soft (120-400~eV) and hard (400-1000~eV) band (see Sec. 3), hence we list the total number of counts. The count numbers for the HRC observations are comparable to those listed in KvK05, but are a factor 100 less than those listed in vK07 (probably due to missprint in vK07) although the data handling in all three works (KvK05, vK07 and this work) is comparable. The count numbers of all other observations are in good agreement. For details on ROSAT and \xmm\ observations we refer to \citealt{2004MNRAS.351.1099C} and \citealt{2009A&A...498..811H} (and Sec. 2, 3 this work), respectively.
}

\longtab{2}{
\begin{longtable}{rl|ll|ll}
\caption{Individual periods and TOAs of the observations of \rxj\ as listed in Tab.~\ref{chandraobs}.}\label{chandraobs2}\\
\hline
Obs. Id.	& Instrument/setup 	 &    period           &          period               & TOA   &     TOA          \\
        	&                  	 &    (120-400~eV)        &         (400-1000~eV)             & (120-400~eV)     & (400-1000~eV) \\     
		      &                    & 	[s]		  &		[s]		      & [days]&     [days]          \\
\hline
\endfirsthead
\caption[]{-- Continued. --}\\\hline
Obs. Id.	& Instrument/setup 	 &    period           &          period ]               & TOA   &     TOA          \\
        	&                  	 &    (120-400~eV)        &         (400-1000~eV)             & (120-400~eV)      & (400-1000~eV) \\     
		      &                    & 	[s]		  &			[s]	      & [days]&     [days]          \\
\hline
\endhead
\hline
\endfoot
\hline
\endlastfoot

rp300338n00 	& PSPC         	 & \multicolumn{2}{c|}{8.39120(44)}				& 49,257.2547153(16)  &   49,257.2547298(21)  \\
rh300508n00 	& HRI          	 & \multicolumn{2}{c|}{8.3852(63)}     				& 50,198.6873383(25)  &   50,198.6873509(35) \\ 
rh180100n00 	& HRI          	 & \multicolumn{2}{c|}{8.3443(11)}     				& 50,210.5562791(18)  &   50,210.5562812(17)  \\
rh300508n01 	& HRI          	 & \multicolumn{2}{c|}{8.4902(98)}     				& 50,353.9975633(27)  &   50,353.9975702(20)  \\ 
rh400884n00 	& HRI          	 & \multicolumn{2}{c|}{8.391130(50)}   				& 50,391.3004644(14)  &   50,391.3004729(18)  \\
h400944n00 	& HRI          	 & \multicolumn{2}{c|}{8.3921(10)}     				  & 50,925.6878393(11)  &   50,925.6878472(21)  \\   
368$^{1}$     	& HRC-S/LETG    	 & \multicolumn{2}{c|}{8.39063(49)}    				& 51,577.0395641(21)  &   51,577.0395693(16)  \\
745$^{1}$     	& HRC-S/LETG    	 & \multicolumn{2}{c|}{}					&  &    \\
369$^{1}$     	& HRC-S/LETG    	 & \multicolumn{2}{c|}{} 					&  &    \\     
0078 S3		& EPIC-pn/FF thin	 &       		8.391113(18)	& 8.391085(35)  	& 51,677.44324000(30) &   51,677.44323979(69) \\  
     S1		& EPIC-MOS1/FF thin	 &       		8.391090(45)	& 8.391090(55)  	& 51,677.47179274(83) &   51,677.47179286(78) \\  
     S2		& EPIC-MOS2/SW thin 	 &    		   	8.391113(23)	& 8.391033(53)  	& 51,677.47179234(54) &   51,677.4717923(13)  \\
0175 S3		& EPIC-pn/FF med & 			8.391268(58)    & 8.39124(12)  	 	& 51,869.95710586(60) &   51,869.95710803(84) \\ 
     S7		& EPIC-MOS1/LW med & 			8.39122(22)     & 8.39107(32)  	 	& 51,869.9949836(13)  &   51,869.9949874(27)  \\  
2774$^{2}$   	& ACIS-CC         	 &			8.391093(13)	& 8.391133(13)   	& 52,248.6767290(15)  &   52,248.67672069(69) \\ 
2773$^{2}$   	& ACIS-CC          	 &					& 		   	&  &    \\ 
2771$^{2}$ 	& ACIS-CC         	 &					& 		   	&  &    \\ 
2772$^{2}$ 	& ACIS-CC         	 &					& 		   	&  &    \\ 
0533 S3 & EPIC-pn/FF thin	 & 			8.391143(43)	& 8.391098(73)  & 52,584.92605120(69) &   52,584.9260530(10)  \\
     S1	& EPIC-MOS1/FF thin	 & 			8.39136(12) 	& 8.39088(16)		& 52,584.91993465(11) &   52,584.9200294(15)  \\   
     S2	& EPIC-MOS2/FF thin	 & 			8.39122(12) 	& 8.39104(20)		& 52,584.91993465(95) &   52,584.9199325(16)  \\
0534 S3	& EPIC-pn/FF thin	 & 			8.391225(40)	& 8.391317(63)  & 52,587.00129529(39) &   52,587.00129506(97) \\ 
     S1	& EPIC-MOS1/FF thin	 & 			8.39119(11) 	& 8.39116(18)		& 52,586.9952779(18)  &   52,586.9952766(29)  \\
     S2	& EPIC-MOS2/FF thin	 & 			8.39137(13) 	& 8.39137(23)		& 52,586.99527264(93) &   52,586.9952735(14)  \\
0622 U2	& EPIC-pn/SW thick	 & 			8.391133(18)	& 8.391138(23)  & 52,761.99514720(82) &   52,761.99514812(64) \\
     S5	& EPIC-MOS1/FF med & 			8.39130(16) 	& 8.39100(20)		& 52,762.2413478(13)  &   52,762.2413532(20)  \\
     S3	& EPIC-MOS1/FF thin	 & 			8.39101(13) 	& 8.39100(17)		& 52,761.8668566(19)  &   52,761.8668516(28)  \\
     S6	& EPIC-MOS2/FF med & 			8.39100(15) 	& 8.39126(15)		& 52,762.2413439(17)  &   52,762.24144068(14) \\
     S4	& EPIC-MOS2/FF thin	 & 			8.390950(95)	& 8.39118(21)		& 52,761.8669520(11)  &   52,761.8669491(20)  \\
0711 S8	& EPIC-pn/SW med & 			8.391082(48)	& 8.391188(82)  & 52,940.11635982(97) &   52,940.11627407(69) \\  
     S3	& EPIC-MOS1/FF thin	 & 			8.39125(31) 	& 8.39077(52)		& 52,939.99012307(84) &   52,939.9900211(20)  \\   
     S5	& EPIC-MOS1/LW thin	 & 			8.39087(24) 	& 8.39059(41)		& 52,940.1679353(12)  &   52,940.1679433(29)  \\ 
     S4	& EPIC-MOS2/FF thin	 & 			8.39169(32) 	& 8.39051(49)		& 52,939.9900070(17)  &   52,939.9900222(24)  \\  
     S6	& EPIC-MOS2/LW thin	 & 			8.39091(22) 	& 8.39147(26)		& 52,940.1679318(13)  &   52,940.1679407(14)  \\
4666$^{3}$      & ACIS-CC          	 & 			8.3911142(23)	& 8.3911216(28)& 53,016.68131016(91) &   53,016.6813159(12)  \\ 	
4667$^{3}$      & ACIS-CC         	 & 					& 		   	&  &    \\
4668$^{3}$      & ACIS-CC          	 & 					& 		   	&  &    \\
4669$^{3}$      & ACIS-CC               & 					& 		   	&  &    \\
5305$^{4}$	& HRC-S/LETG            & \multicolumn{2}{c|}{8.39109(25)}				& 53,062.4157092(15)  &   53,062.4157091(15)  \\
0815 S1 & EPIC-pn/FF thin   & 			8.391105(35)    & 8.391143(58)	      & 53,147.68119387(43) &   53,147.6812041(15)  \\ 
     S2	& EPIC-MOS1/FF thin   & 			8.391108(68)    & 8.39111(15)	        & 53,147.6883801(11)  &   53,147.6883883(31)  \\
     S3	& EPIC-MOS2/FF thin   & 			8.391150(60)    & 8.39120(13)	        & 53,147.68838022(93) &   53,147.6883911(21)  \\
4670$^{5}$      & ACIS-CC          	 & 			8.3911146(15)	& 8.3911165(16)  	& 53,230.5756209(11)  &   53,230.5756243(19)  \\
4671$^{5}$      & ACIS-CC               &					&			&  &    \\
4672$^{5}$      & ACIS-CC               &					&			&  &    \\
4673$^{5}$      & ACIS-CC               &					&			&  &    \\
5581$^{6}$    	& HRC-S/LETG            & \multicolumn{2}{c|}{8.391010(95)}				& 53,393.6674769(19)  &   53,393.6674775(23)  \\
0986 S3	& EPIC-pn/FF thin	 &			8.391138(23)	& 8.391140(30)  	& 53,488.67561302(44) &   53,488.67561797(84) \\
     S1	& EPIC-MOS1/SW thin 	 &			8.391095(45)	& 8.391008(48)  	& 53,488.6695937(14)  &   53,488.6695969(13)  \\
     S2	& EPIC-MOS2/SW thin 	 &			8.391183(38)	& 8.391018(53)  	& 53,488.6696892(13)  &   53,488.6696931(14)  \\
5582$^{7}$  	& HRC-S/LETG    	 & \multicolumn{2}{c|}{8.391138(63)}				& 53,522.9398395(35)  &   53,522.9398385(29)  \\
6364$^{8}$	  & HRC-S/LETG      & \multicolumn{2}{c|}{8.39129(25)}				&  53,610.0881154(30)  &   53,610.0881131(20)  \\
1060 S3	& EPIC-pn/FF thin	 &			8.391120(20)	& 8.391082(33) 	 	& 53,636.30015746(23) &   53,636.3001616(10)  \\
     S1	& EPIC-MOS1/SW thin 	 &			8.391095(40)	& 8.391130(50) 	 	& 53,636.29413571(70) &   53,636.2941416(13)  \\
     S2	& EPIC-MOS2/SW thin 	 &			8.391103(38)	& 8.391163(58) 	 	& 53,636.29413754(96) &   53,636.29413962(83) \\
6369$^{9}$    	& HRC-S/LETG            & \multicolumn{2}{c|}{8.39110(11)}	       			&  53,652.2601916(17)  &   53,652.2601885(15)  \\ 
7177$^{9}$    	& HRC-S/LETG            &					&			&  &    \\
1086 S3	& EPIC-pn/FF thin	 &			8.391125(25)	& 8.391072(38)	   	& 53,687.17180535(53) &   53,687.17180817(92) \\
     S1	& EPIC-MOS1/SW thin 	 &			8.391003(63)	& 8.391085(80)	   	& 53,687.1657835(12)  &   53,687.16588680(52) \\ 
     S2	& EPIC-MOS2/SW thin 	 &			8.391200(55)	& 8.390998(78)	   	& 53,687.1658825(10)  &   53,687.1658857(12)  \\
7243$^{10}$   	& HRC-S/LETG            & \multicolumn{2}{c|}{8.391115(10)}   	& 53,720.0243035(13)  &   53,720.0243020(12)  \\  
7244$^{10}$   	& HRC-S/LETG            &					&			&  &    \\
7245$^{10}$   	& HRC-S/LETG            &					&			&  &    \\
5584$^{10}$   	& HRC-S/LETG            &					&			&  &    \\
{\it 7251}	& {\it HRC-S/LETG}  & \multicolumn{2}{c|}{{\it 8.39212(88)}} & {\it 53,775.3509131(17)}& {\it 53,775.35091144(12) }\\  
1181 S1	& EPIC-pn/FF thin	 &			8.391090(70)	& 8.39120(11) 	  & 53,877.32857154(71) &   53,877.3285752(11)  \\
     S2	& EPIC-MOS1/SW thin 	 &			8.39137(17) 	& 8.39101(20) 	  & 53,877.3225489(15)  &   53,877.32255272(65) \\
     S3	& EPIC-MOS2/SW thin 	 &			8.39131(13)   & 8.39116(18) 	  & 53,877.3225491(11)  &   53,877.3225524(16)  \\
1265 S1	& EPIC-pn/FF thin	 &			8.391075(65)	& 8.39104(11)  	 	& 54,044.60537610(44) &   54,044.6053791(11)  \\
     S2	& EPIC-MOS1/FF thin	 &			8.39117(16) 	& 8.39074(17)  	 	& 54,044.5990605(12)  &   54,044.5990664(13)  \\
     S3	& EPIC-MOS2/FF thin	 &			8.39099(13) 	& 8.39096(19)  	 	& 54,044.59906220(62) &   54,044.5990665(18)  \\
1356 S1	& EPIC-pn/FF thin	 &			8.391150(65)	& 8.391113(98) 	 	& 54,225.84098865(82) &   54,225.84099190(98) \\
     S2	& EPIC-MOS1/FF thin	 &			8.39107(21) 	& 8.39066(27)  	 	& 54,225.8345794(18)  &   54,225.83457862(99) \\
     S3	& EPIC-MOS2/FF thin	 &			8.39103(23) 	& 8.39096(22)  	 	& 54,225.8346806(13)  &   54,225.8346782(22)  \\
1454 S1	& EPIC-pn/FF thin	 &			8.391130(50)	& 8.391085(85) 	 	& 54,421.36989356(53) &   54,421.36989479(60) \\
     S2	& EPIC-MOS1/FF thin	 &			8.39090(16) 	& 8.39109(17)  	 	& 54,421.3635797(17)  &   54,421.3635840(15)  \\
     S3 & EPIC-MOS2/FF thin	 &			8.39131(22) 	& 8.39127(22)  	 	& 54,421.3635875(20)  &   54,421.3635849(18)  \\
10861$^{11}$	       & HRC-S/LETG            & \multicolumn{2}{c|}{8.39129(20)}				& 54,863.7916775(17)  &   54,863.7916751(12)  \\ 
10700$^{11}$   	     & HRC-S/LETG            &					&			&  &    \\
1700 S3	& EPIC-pn/FF thin	 &			8.39108(99)	& 8.3912(12)		& 54,911.7856154(16)  &   54,911.78046975(81) \\
     U2	& EPIC-MOS1/FF thin	 &			8.39097(26)	& 8.39136(29)		& 54,911.7440458(30)  &   54,911.7440479(17)  \\
     U2	& EPIC-MOS2/FF thin	 &			8.39103(25)	& 8.39118(35)		& 54,911.7440524(13)  &   54,911.7440483(15)  \\
10701$^{12}$   	    & HRC-S/LETG            &					& 8.39119(20)    	& 55,085.6684206(14)  &   55,085.6684187(15)  \\
1792 S3	& EPIC-pn/FF thin	 &			8.39102(12)	& 8.39101(15)		& 55,096.30446151(69) &   55,096.3044632(11)  \\
     S1	& EPIC-MOS1/SW thin 	 &			8.39097(20)	& 8.39120(20)		& 55,096.2966932(13)  &   55,096.2966943(13)  \\
     S2	& EPIC-MOS2/SW thin 	 &			8.39104(15)	& 8.39102(24)		& 55,096.2967896(14)  &   55,096.2967885(11)  \\
					        							 
\hline					        										      
\end{longtable}
\noindent Note: The individual periods for ROSAT and Chandra HRC were not divided into soft (120-400~eV) and hard (400-1000~eV) band (see Sec. 3), hence we list the individual periods derived from all counts. We list the TOAs (definition equal to that in KvK05 and vK07) of the soft band (for the ``all data" solution of the soft band, see Tab.~\ref{zusammenfassung}) and of the hard band (for the ``all data" solution of the hard band, see Tab.~\ref{zusammenfassung}). The errors of the periods \citep{2002AJ....124.1788R} and the TOAs (derived from the fitted light curves, see text) are given in parenthesis and denote 1$\sigma$ confidence level.
}

\end{document}